# Controlled Growth and Disorder Mitigation of $GaAs_{1-x}Sb_x$/InAs Core-Shell Quantum Wires

Ran Zhuo[†,⊥,#], Huading Song[‡,#], Yanlin Wu[†,⊥,#], Wenlong Yang[†], Haiyan Shi[†], Xiyu Hou[†,⊥], Sijun Chen[§], Zhichuan Wang[||], Donghai Li[†,⊥], Jinjin Bai[†], Runan Shang[‡], Hao Zhang[§,||]*, Dong Pan[†,⊥,]*

[†]*State Key Laboratory of Semiconductor Physics and Chip Technologies, Institute of Semiconductors, Chinese Academy of Sciences, Beijing 100083, China*

[⊥]*Center of Materials Science and Optoelectronics Engineering, University of Chinese Academy of Sciences, Beijing 100049, China*

[‡]*Beijing Academy of Quantum Information Sciences, Beijing 100193, China*

[§]*College of Semiconductors, Southern University of Science and Technology, Shenzhen 518055, China*

[||]*Quantum Science Center of Guangdong-Hong Kong-Macao Greater Bay Area, Shenzhen 518045, China*

## ABSTRACT

We report the growth and transport properties of $GaAs_{1-x}Sb_x$/InAs core-shell nanowires. Compared with the well-studied GaAs-InAs system, incorporating Sb substantially reduces the lattice mismatch between the core and shell. Accordingly, both the nanowire crystal quality and device performance improved significantly. Increasing $x$ from 0.15 to 0.35, the density of crystalline defects, such as stacking faults in the InAs shell, decreases by half. Moreover, the field-effect mobility increases from 3850 $cm^2/Vs$ to 4800 $cm^2/Vs$, consistent with the reduction of the lattice mismatch. The incorporation of Sb makes the nanowire easier to pinch off via the gate voltage, enhancing the electric field tunability over the nanowire. Magneto-conductance measurements reveal Aharonov-Bohm-like oscillations, suggesting a phase-coherent tubular InAs shell. Our results identify lattice mismatch as a critical disorder source in core-shell nanowires and provide a practical route toward low-disorder core-shell quantum devices.

*To whom correspondence should be addressed. E-mails: zhanghao@sustech.edu.cn; pandong@semi.ac.cn

#These authors contributed equally to this work.

InAs nanowires are widely studied as a promising materials platform for various quantum devices. For example, hybrid InAs-superconductor nanowires are among the leading materials platforms for realizing topological superconductivity[1-10]. A particularly interesting architecture is the core-shell geometry, in which a wide-band-gap core is overgrown with an InAs shell. This well-defined ring geometry forms a tubular conductor that can be phase biased by threading magnetic flux through the core. Indeed, Aharonov-Bohm oscillations in magnetoconductance have been reported in GaAs/InAs core-shell nanowires, demonstrating phase-coherent transport through a tubular conduction channel[11-15]. More recently, $h/2e$-periodic supercurrent oscillations have also been observed in Josephson junctions based on these nanowires. The full proximitization of the InAs shell is relevant to the non-trivial topology of the Andreev spectrum[16-18]. Another related platform is provided by full-shell InAs-Al nanowires[19,20]. In these devices, magnetic flux provides an additional tuning parameter for exploring possible topological superconductivity.

Despite these promising aspects, a fundamental challenge of the GaAs/InAs system is its large lattice mismatch. The lattice constants of GaAs and InAs are 5.65 Å and 6.05 Å, respectively. This 7% mismatch inevitably introduces a high density of misfit dislocations, stacking faults, and other crystalline defects at the GaAs/InAs interface. Such disorder severely limits the electronic mobility and device tunability. In fact, GaAs/InAs core-shell nanowires are difficult to pinch off using electrostatic gates, limiting their use in quantum devices[21]. Eliminating or mitigating this disorder is therefore essential[22].

In this work, we demonstrate a promising strategy to mitigate the adverse effects of lattice mismatch. We replaced the GaAs core with $GaAs_{1-x}Sb_x$. Because the lattice constant of GaSb (6.09 Å) is nearly identical to that of InAs, increasing the Sb content $x$ from 0 to 1 continuously tunes the lattice constant of $GaAs_{1-x}Sb_x$ toward that of InAs. As a result, $GaAs_{0.85}Sb_{0.15}$/InAs core-shell nanowires can be pinched off using electrostatic gates and yield an average mobility of 3850 $cm^2$/Vs. In addition, the averaged mobility of $GaAs_{0.65}Sb_{0.35}$/InAs nanowires increases to 4800 $cm^2$/Vs. This 25% mobility enhancement is consistent with the reduced lattice mismatch at higher Sb content, highlighting the pivotal role of lattice matching in suppressing disorder. Moreover, magneto-conductance measurements reveal Aharonov-Bohm-like oscillations. The oscillation period yields an interference area matching the nanowire cross section, consistent with phase-coherent transport through the tubular InAs channel. Our results establish lattice mismatch as a key source of disorder in core-shell nanowires and, more importantly, provide a practical route to mitigating it. The improved crystal quality and mobility pave the way for applications in controllable quantum devices such as topological superconductivity.

**Nanowire growth**

All $GaAs_{1-x}Sb_x$/InAs core-shell nanowires were grown in an MBE system (VG V80H) using commercial n-type Si (111) wafers as substrates. Prior to nanowire growth, the Si

(111) wafers underwent a two-step surface-preparation procedure. The native oxide was first removed by immersing the substrates in an 8% hydrofluoric acid solution for approximately 40 s. The substrates were then immediately subjected to a 15 s treatment in a 4:1 (by volume) mixture of 98% $H_2SO_4$ and 30% $H_2O_2$ to form a fresh, thin oxide layer on the cleaned surface.

Three sets of self-catalyzed $GaAs_{1-x}Sb_x$ nanowires with different Sb compositions were initially grown. The growth conditions for the $GaAs_{1-x}Sb_x$ nanowire cores were as follows: the Ga and As beam equivalent pressures (BEPs) were fixed at $3.0 \times 10^{-7}$ mbar and $2.0 \times 10^{-6}$ mbar, respectively, while the Sb BEP values were set to $4.0 \times 10^{-7}$ mbar, $8.0 \times 10^{-7}$ mbar, and $1.1 \times 10^{-6}$ mbar, respectively. The growth temperature and duration were 660 °C and 30 min, respectively. Immediately after growing the $GaAs_{1-x}Sb_x$ core nanowires, an InAs shell was epitaxially grown *in situ*. The InAs shells were grown at 500 °C with an As/In BEP ratio of 5.0 (In BEP = $4.0 \times 10^{-7}$ mbar, As BEP = $2.0 \times 10^{-6}$ mbar) for 45 min. Using this procedure, we prepared three sets of $GaAs_{1-x}Sb_x$/InAs core-shell nanowire samples with Sb contents of $x = 0.15$, 0.30, and 0.35.

**Microstructural analysis**

Nanowire morphology was characterized by scanning electron microscopy (SEM; FEI NanoSEM 650, 10 kV). Crystal structure and elemental distributions were examined using transmission electron microscopy (TEM; JEM-F200, 200 kV) and energy-dispersive X-ray spectroscopy (EDS). For TEM imaging, nanowires were released from the substrate by ultrasonication in ethanol, and the resulting suspension was drop-cast onto carbon-coated grids. Cross-sectional lamellae were prepared by depositing a protective Pt layer over the target region and then performing ion-beam cutting and lift-out. The extracted lamellae were then transferred to dedicated TEM holders. Final thinning and cleaning with low-voltage ion milling produced electron-transparent areas suitable for detailed TEM analysis.

**Electrical characterization**

After growth, the $GaAs_{1-x}Sb_x$/InAs nanowires were transferred to p-doped $Si/SiO_2$ substrates for device fabrication. Electron-beam lithography was used to pattern electrical contacts. Prior to metal evaporation, we performed Ar plasma etching in the load-lock chamber at 50 W and 0.05 Torr for 80 s to obtain ohmic contacts. This step may damage the InAs shell underneath the contacts and affect mobility. The Ti/Au (8/100 nm) contact electrodes were deposited by electron beam evaporation. The electrical characterization was performed in a dilution refrigerator with a base temperature of approximately 15 mK. The DC bias ($V$) and AC lock-in excitation ($dV$) passed through a 1000:1 voltage divider and then through a series of filters (a copper-powder filter and a RC filter at the mixing chamber) before being applied to the device. The device currents ($I$ and $dI$) passed through the same filter chain to a preamplifier (gain: 1 μA/V) and were recorded by a lock-in amplifier and a DC voltmeter.

**Epitaxial Growth of InAs Shells on Ternary $GaAs_{1-x}Sb_x$ Nanowires with Different Sb Contents**

As discussed above, the lattice mismatch between ternary $GaAs_{1-x}Sb_x$ and InAs decreases with increasing Sb content. This compositional dependence provides a route to improving the crystalline quality of the InAs shell by tailoring the Sb content in the $GaAs_{1-x}Sb_x$ core. Accordingly, we grew three sets of $GaAs_{1-x}Sb_x$/InAs core-shell nanowire samples with different Sb contents. Figure 1a schematically illustrates the growth procedure. The ternary $GaAs_{1-x}Sb_x$ cores (highlighted in red) were grown via Ga-self-catalyzed growth. Specifically, Ga droplets first formed on the Si (111) substrate and subsequently catalyzed the axial growth of $GaAs_{1-x}Sb_x$ nanowire cores; the Sb content $x$ was tuned by adjusting the Sb beam flux. Subsequent EDS measurements determined the Sb contents $x$ of the three samples to be 0.15, 0.30, and 0.35, respectively. After core growth, the Ga droplet at the nanowire tip was removed by exposure to an $As_4$ flux. Finally, the InAs shells (shown in blue) were epitaxially deposited *in situ* on the core sidewalls. It is well established that the quality of ternary $GaAs_{1-x}Sb_x$ nanowires directly dictates the crystalline quality of the subsequently grown InAs shell. Consequently, in the present study, high-quality $GaAs_{1-x}Sb_x$ nanowires were employed as core templates for InAs shell growth[23], thereby effectively precluding any degradation of the shell quality that could otherwise originate from defects in the core. Figures 1b-d present side-view SEM images of the three samples. These images confirm the successful growth of $GaAs_{1-x}Sb_x$/InAs core-shell nanowires on Si (111) substrates; all nanowires are vertically aligned and grow along the ⟨111⟩ direction. Moreover, the core-shell nanowires exhibit a distinct pencil-like morphology, with smooth surfaces and nearly uniform diameters; only slight variations are observed near their tips. Previous studies indicate that incomplete shell growth typically manifests as discontinuous islands or irregular surface features[24]. The SEM observations therefore show that continuous InAs shells were epitaxially grown on the sidewalls of the ternary $GaAs_{1-x}Sb_x$ nanowires. Across the Sb-content range $x$ = 0.15-0.35, the InAs shells remained continuous and smooth. As shown in Figs. 1e-g, tilted-view (25°) SEM images further confirm the smooth and continuous morphology of the $GaAs_{1-x}Sb_x$/InAs core-shell nanowires. Together, these results show that well-faceted, high-quality InAs shells can be epitaxially grown on ternary $GaAs_{1-x}Sb_x$ nanowire cores over a broad range of Sb compositions. For more information on the diameter, length, and shell thickness of the $GaAs_{1-x}Sb_x$/InAs core-shell nanowires, please refer to Figs. S1 to S3 in Supporting Information.

**Microstructure and Composition of InAs Shells Epitaxially Grown on Ternary $GaAs_{1-x}Sb_x$ Nanowires with Different Sb Contents**

Detailed TEM measurements were used to evaluate the crystalline quality of InAs shells grown on ternary $GaAs_{1-x}Sb_x$ nanowires with different Sb contents. Figures 2a and 2b show low-magnification TEM images of representative nanowires with $x$ = 0.15 and 0.35, respectively. In both samples, the InAs shells uniformly encapsulate the $GaAs_{0.85}Sb_{0.15}$ and $GaAs_{0.65}Sb_{0.35}$ cores, and the diameter decreases only near the nanowire tip, consistent with the SEM observations. Figures 2d,f,h,and j show high-resolution TEM images acquired from different areas of the $GaAs_{0.85}Sb_{0.15}$/InAs nanowire ($x$ = 0.15), whereas Figs. 2e,g,i,and k show the corresponding fast Fourier transform (FFT) patterns. These data show that the InAs shell adopts a pure zinc-blende crystal structure. Nonetheless, numerous defects, including twins and stacking faults, are evident in Figs. 2f,g,j,and k. The stacking-fault density is approximately 10 $\mu m^{-1}$, likely because of the relatively large lattice mismatch at low Sb content. The FFT pattern in Fig. 2e exhibits two distinct sets of diffraction spots, confirming epitaxial alignment between the $GaAs_{1-x}Sb_x$ core and the InAs shell and the formation of a well-defined core-shell heterostructure. Figures 2l,n,p,and r show high-resolution TEM images acquired from distinct regions of the $GaAs_{0.65}Sb_{0.35}$/InAs nanowire ($x$ = 0.35), whereas Figs. 2m,o,q,and s show the corresponding FFT patterns. The images and FFT patterns from different regions show that the InAs shell retains a pure zinc-blende single-crystal structure. More importantly, the crystalline quality improves substantially at the higher Sb content, with only occasional stacking faults observed (marked by the yellow arrow in Fig. 2n). Statistical analysis yields an InAs-shell defect density of approximately 5 $\mu m^{-1}$. As shown in Figs. 2m,o,q,and s, the corresponding FFT patterns further confirm a consistent crystal orientation throughout the InAs shells, with no evidence of twin defects. Notably, the FFT patterns show nearly overlapping diffraction spots from the core and shell regions, consistent with the smaller lattice mismatch and more closely matched lattice constants between high-Sb $GaAs_{1-x}Sb_x$ and InAs. These observations demonstrate that increasing the Sb content improves the crystalline quality of the InAs shells. In addition, $GaAs_{0.70}Sb_{0.30}$/InAs core-shell nanowires (see Fig. S4 in Supporting Information) were examined and likewise exhibited zinc-blende single-crystal InAs shells with a defect density of ~6 μm$^{-1}$, further confirming the beneficial effect of increasing the Sb content. For more information on the statistical analysis of the defect density in the $GaAs_{1-x}Sb_x$/InAs core-shell nanowires, please refer to Figs. S5 and S6 in Supporting Information.

To investigate the microstructure and confirm that the obtained samples are indeed $GaAs_{1-x}Sb_x$/InAs core-shell nanowires, cross-sectional specimens were prepared for detailed TEM and EDS characterization. Figure 3 presents the typical crystal structure and chemical composition of a cross-sectional sample of the $GaAs_{0.65}Sb_{0.35}$/InAs core-shell nanowire. The low-magnification cross-sectional TEM image in Fig. 3a reveals a regular hexagonal cross section bounded by six {110} facets. A well-defined core-shell structure with a continuous, uniform InAs shell encapsulating the $GaAs_{0.65}Sb_{0.35}$ core is evident. As shown in Figs. 3b-d, cross-sectional high-resolution TEM images confirm that both the $GaAs_{0.65}Sb_{0.35}$ core and the InAs shell have high crystalline quality and a pure zinc-blende structure, with no significant defects observed. The contrast variation in Fig. 3d approximately delineates the core-shell interface. EDS elemental mapping (Figs. 3e-g) and line scans (Fig. 3h) reveal a sharp heterointerface between the $GaAs_{0.65}Sb_{0.35}$ core and the InAs shell, confirming high-quality epitaxy. Here, we presented only the Ga and In EDS maps because Ga is exclusively present in the core and In exclusively in the shell, and these two elements alone therefore provide unambiguous evidence for the formation of the core-shell geometry. From the growth perspective, we can additionally argue that Sb segregation, interface intermixing, and Sb depletion from the core are unlikely. The $GaAs_{1-x}Sb_x$ cores were first grown at a high temperature (660 °C) for 30 min, after which the growth temperature was lowered to the InAs shell growth temperature (500 °C) for the *in-situ* deposition of the InAs shell. On the basis of growth thermodynamics and kinetics, the Sb atoms are strongly bonded within the $GaAs_{1-x}Sb_x$ core lattice at the lower shell-growth temperature, and the low thermal budget of the shell growth provides neither the driving force nor the atomic mobility required for significant Sb segregation or intermixing. In addition, the InAs shell thickness is uniform. Based on high-resolution TEM and EDS analyses, the core diameter and shell thickness of this nanowire are estimated to be approximately 82 nm and 22 nm, respectively. Notably, the diameters of individual $GaAs_{1-x}Sb_x$ nanowires on the same substrate exhibit a distribution, which is attributable to Ostwald ripening. Subsequently, to obtain thickness information of the specimens used for transport measurements, cross-sectional TEM imaging has been carried out on all nanowires after completion of the transport experiments, which further confirms that increasing the Sb content improves the crystalline quality of the InAs shell (see Figs. S7 and S8 in the Supporting Information for details).

**Transport Properties of InAs Nanowires Epitaxially Grown on Ternary $GaAs_{1-x}Sb_x$ Nanowires with Different Sb Contents**

Figure 4a shows a field-effect-transistor fabricated from a $GaAs_{1-x}Sb_x$/InAs nanowire with $x$ = 0.35 (Device A). The lower panel shows a schematic of the wire cross-section.

The $GaAs_{0.65}Sb_{0.35}$ core has a radius of ~39.5 nm and the InAs shell has an outer radius of 65.2 nm. The back gate consists of p-doped Si, separated from the nanowire by a 307-nm-thick $SiO_2$ dielectric layer. All measurements were two-terminal and performed in a dilution refrigerator at its base temperature (about 15 mK). The *I*-*V* curves in Fig. 4b exhibit clear linearity at various back-gate voltages $V_{BG}$. We then fixed the bias voltage $V$ around 9 mV and measured the DC current $I$. Figure 4c shows the DC conductance $G = I/V$ as a function of $V_{BG}$. The sweeping direction of $V_{BG}$ is from negative to positive (upward). The series resistance contributed by filters has been extracted from the two-terminal circuit. The red and blue curves correspond to devices A and B, respectively; device B has a $GaAs_{0.85}Sb_{0.15}$ core ($x = 0.15$). The dashed lines are fits used to extract the mobility, based on the formula: $\frac{1}{G} = R_s + \frac{L^2}{\mu C(V_{BG} - V_{pf})}$. $R_s$ denotes the contact resistance. $L$ is the channel length (contact spacing), measured from the SEM. $\mu$ is the mobility, and $V_{pf}$ is the pinch-off voltage. $C$ is the gate capacitance estimated from the device geometry (see the schematic in Fig. 4a). Following Ref. 25, we used a finite element model that incorporates a 20% reduction of the classical capacitance to account for quantum confinement. For each Sb composition ($x = 0.15$ or 0.35), we measured two devices, denoted as A and B. Figure 4d shows the up-sweep pinch-off curves of all four devices. Their mobilities are summarized in Fig. 4e. The average device mobilities measured in the same gate-sweeping direction are displayed in blue ($x = 0.15$) and red ($x = 0.35$). The average mobility for devices with $x = 0.35$ is 4800 $cm^2$/Vs, which is approximately 25% higher than that for $x = 0.15$ devices (3850 $cm^2/Vs$). Standard deviations are depicted by the error bars. As a comparison, literature reported mobility of GaAs/InAs core-shell nanowires ($x$=0) to be generally below 500 $cm^2/Vs$ [21], significantly smaller than that of both cases here. The lattice constant of GaAs, $GaAs_{0.85}Sb_{0.15}$, and $GaAs_{0.65}Sb_{0.35}$ are 5.65 Å, 5.72 Å, and 5.80 Å, respectively. The lattice mismatch relative to InAs (6.05 Å) is 6.8%, 5.6%, and 4.2%, correspondingly. Figure S9 and Table S1 in Supplementary Information provide detailed information on all devices. For both sweep directions, the mobilities of InAs nanowires on a $GaAs_{0.65}Sb_{0.35}$ core (red bars) are higher than those of nanowires on a $GaAs_{0.85}Sb_{0.15}$ core (blue bars). We attribute this enhancement to the higher Sb content ($x$), which reduces the lattice mismatch between the core and the shell. In addition, the threshold voltages are more negative for the $x$ =0.15 case (blue), indicating that the devices are harder to pinch off when the core-shell lattice mismatch is larger. The threshold voltage ($V_{th}$) in Fig. 4f is defined as the $V_{BG}$ value at which G is 0.01× $2e^2/h$.

We next investigate the magnetoconductance of device A, as shown in Fig. 5. The magnetic field ($B$) was applied along the nanowire axis. Figure 5a shows the differential conductance $dI/dV$ measured by a lock-in amplifier as a function of $B$. $V$ was fixed at zero. $V_{BG}$ = 33.5 V. Quasi-periodic oscillations are observed. As a guide to the eye, vertical dashed lines are plotted at 0.4-T intervals. The grey curve shows a background obtained by smoothing the data over a 0.4-T window. Subtracting this background highlights the oscillations $\Delta G$, as shown in Fig. 5b. The FFT of $\Delta G$ (Fig. 5c) shows a dominant peak near $(0.4\,T)^{-1}$ = 2.5 $T^{-1}$ (the green dashed line). Using the relation $\Delta B \cdot A = h/e$, a field period of 0.4 T in $\Delta B$ corresponds to an Aharonov-Bohm interference area $A$ = 10340 $nm^2$. Here, $h$ is Planck's constant. This area corresponds to a hexagonal cross-section with a radius of 63 nm. Based on the dimensions of device A shown in Fig. 4a, this radius is close to the outer radius of InAs shell (65.2 nm) measured using TEM. These oscillations are therefore consistent with an Aharonov-Bohm-like origin, with the interfering electrons located near the outer surface of the InAs shell. This interpretation is consistent with the band structure of these core-shell devices, in which band bending produces surface accumulation near the outer surface of InAs[16].

Figure 5d shows additional oscillation curves measured at different $V_{BG}$ values. Their averaged FFT amplitudes are shown in Fig. 5e (red curve), with the pink shaded region indicating the standard deviation. The dominant peak at 2.5 $T^{-1}$ (green dashed line) persists, suggesting that the Aharonov-Bohm-like interference is robust against changes in gate voltage. A second peak near 4 $T^{-1}$ is also visible. Its frequency is not exactly twice the 2.5 $T^{-1}$ fundamental but a factor of 1.6. If the peak is tentatively attributed to an $h/2e$ contribution, the relation $\Delta B \cdot A = h/2e$ gives an inferred area $A$ = 8270 $nm^2$, corresponding to a hexagonal cross-section with a radius of 56.4 nm. This radius lies between the core radius of 39.5 nm and the outer shell radius of 65.2 nm, near the midpoint of the InAs shell. We note that the oscillation periodicity is less well defined than in nanowire networks[26] with a fixed interference loop area. Here, the finite InAs-shell thickness permits a range of loop areas and therefore broadens the Aharonov-Bohm periodicity. Figure S10 in Supplementary Information shows the magnetoconductance of device A for $B$ perpendicular to the substrate; these data also exhibit quantum interference oscillations. The loop areas inferred from the oscillation periods differ from the cross-section area of device A, indicating that the relevant interference loops likely lie in planes containing the nanowire axis rather than in the cross-sectional plane. Aharonov-Bohm-like oscillations are also observed in another $x$ = 0.35 device (device C), as shown in Fig. S11.

In summary, we have demonstrated the controlled growth of $GaAs_{1-x}Sb_x$/InAs core-shell nanowires using MBE. The incorporation of Sb reduces the lattice mismatch between the core and the shell. Consequently, high-quality InAs shells can be realized with fewer crystalline defects and misfit dislocations than shells grown on GaAs cores. As the Sb content ($x$) increases, the defect density decreases and the electron mobility increases, highlighting the critical role of lattice mismatch in core-shell nanowires. Magnetoconductance measurements reveal Aharonov-Bohm-like oscillations, consistent with phase-coherent transport through a tubular InAs shell. Our results demonstrate a practical strategy for mitigating disorder in core-shell nanowires and may support future low-disorder semiconductor-superconductor hybrid devices designed to explore Majorana physics.

## ASSOCIATED CONTENT

Statistical information on the diameter, length, and shell thickness of the $GaAs_{1-x}Sb_x$/InAs core-shell nanowires. Microstructural characterization and defect density analysis of $GaAs_{0.70}Sb_{0.30}$/InAs core-shell nanowires. Statistical analysis of the defect density in the $GaAs_{1-x}Sb_x$/InAs core-shell nanowires. Additional cross-sectional TEM characterization on the lower-Sb samples. Additional transport data of InAs nanowires epitaxially grown on ternary $GaAs_{1-x}Sb_x$ nanowires with different Sb contents.

## AUTHOR INFORMATION

**Corresponding Author**

*E-mails: zhanghao@sustech.edu.cn; pandong@semi.ac.cn;

**Notes**

The authors declare that they have no competing financial interests.

## ACKNOWLEDGEMENTS

[#]These authors contributed equally to this work. This work was supported by the National Natural Science Foundation of China (Grant Numbers 12374459, 61974138 and 92065106), the Quantum Science and Technology-National Science and Technology Major Project (Grant Number 2021ZD0302400), the National Key Research and Development Program of China (Grant Number 2025YFA1411400), and

the Strategic Priority Research Program of Chinese Academy of Sciences (Grant Number XDB0460000). D. P. also acknowledges the support from the State Key Laboratory of Micro-nano Engineering Science (Grant Number MES202601) and Youth Innovation Promotion Association, Chinese Academy of Sciences (Grant Numbers 2017156 and Y2021043). H. S. acknowledges Beijing Natural Science Foundation (Grant No.1252037).

**Conflict of Interest**

The authors declare no competing financial interests.

## REFERENCES

(1) Lutchyn, R. M.; Sau, J. D.; Das Sarma, S. Majorana Fermions and a Topological Phase Transition in Semiconductor-Superconductor Heterostructures. *Phys. Rev. Lett.* **2010**, *105*, 077001.

(2) Oreg, Y.; Refael, G.; Von Oppen, F. Helical Liquids and Majorana Bound States in Quantum Wires. *Phys. Rev. Lett.* **2010**, *105*, 177002.

(3) Mourik, V.; Zuo, K.; Frolov, S. M.; Plissard, S. R.; Bakkers, E. P.; Kouwenhoven, L. P. Signatures of Majorana Fermions in Hybrid Superconductor-Semiconductor Nanowire Devices. *Science* **2012**, *336*, 1003–1007.

(4) Deng, M. T.; Yu, C. L.; Huang, G. Y.; Larsson, M.; Caroff, P.; Xu, H. Q. Anomalous Zero-Bias Conductance Peak in a Nb–InSb Nanowire–Nb Hybrid Device. *Nano Lett.* **2012**, *12*, 6414–6419.

(5) Krogstrup, P.; Ziino, N. L. B.; Chang, W.; Albrecht, S. M.; Madsen, M. H.; Johnson, E.; Nygård J.; Marcus, C. M.; Jespersen, T. S. Epitaxy of Semiconductor–Superconductor Nanowires. *Nat. Mater.* **2015**, *14*, 400–406.

(6) Kang, J. H.; Grivnin, A.; Bor, E.; Reiner, J.; Avraham, N.; Ronen, Y.; Cohen, Y.; Kacman, P.; Shtrikman, H.; Beidenkopf, H. Robust Epitaxial Al Coating of Reclined InAs Nanowires. *Nano Lett.* **2017**, *17*, 7520–7527.

(7) Kanne, T.; Marnauza, M.; Olsteins, D.; Carrad, D. J.; Sestoft, J. E.; de Bruijckere, J.; Zeng, L.; Johnson, E.; Olsson, E.; Grove-Rasmussen, K.; Nygård, J. Epitaxial Pb on InAs Nanowires for Quantum Devices. *Nat. Nanotechnol.* **2021**, *16*, 776–781.

(8) Pan, D.; Song, H.; Zhang, S.; Liu, L.; Wen, L.; Liao, D.; Zhuo, R.; Wang, Z.; Zhang, Z.; Yang, S.; Ying, J.; Miao, W.; Shang, R.; Zhang, H.; Zhao, J. In Situ Epitaxy of Pure Phase Ultra-Thin InAs-Al Nanowires for Quantum Devices. *Chin. Phys. Lett.* **2022**, *39*, 058101.

(9) Wang, Z.; Song, H.; Pan, D.; Zhang, Z.; Miao, W.; Li, R.; Cao, Z.; Zhang, G.; Liu, L.; Wen, L.; Zhuo, R.; Liu, D. E.; He, K.; Shang, R.; Zhao, J.; Zhang, H. Plateau Regions for Zero-Bias Peaks within 5% of the Quantized Conductance Value $2e^2/h$. *Phys. Rev. Lett.* **2022**, *129*, 167702.

(10) Su, H.; Wang, J. Y.; Gao, H.; Luo, Y.; Yan, S.; Wu, X.; Li, G.; Shen, J.; Lu, L.; Pan, D.; Zhao, J.; Zhang, P.; Xu H. Q. Microwave-Assisted Unidirectional Superconductivity in Al-InAs Nanowire-Al Junctions under Magnetic Fields. *Phys. Rev. Lett.* **2024**, *133*, 087001.

(11) Rieger, T.; Luysberg, M.; Schäpers, T.; Grützmacher, D.; Lepsa, M. I. Molecular Beam Epitaxy Growth of GaAs/InAs Core–Shell Nanowires and Fabrication of InAs Nanotubes. *Nano Lett.* **2012**, *12*, 5559–5564.

(12) Rieger, T.; Schäpers, T.; Grützmacher, D.; Lepsa, M. I. Crystal Phase Selective Growth in GaAs/InAs Core–Shell Nanowires. *Cryst. Growth Des.* **2014**, *14*, 1167–1174.

(13) Haas, F.; Wenz, T.; Zellekens, P.; Demarina, N.; Rieger, T.; Lepsa, M.; Grützmacher, D.; Lüth, H.; Schäpers, T. Angle-Dependent Magnetotransport in GaAs/InAs Core/Shell Nanowires. *Sci. Rep.* **2016**, *6*, 24573.

(14) Haas, F.; Zellekens, P.; Lepsa, M.; Rieger, T.; Grützmacher, D.; Lüth, H.; Schäpers, T. Electron Interference in Hall Effect Measurements on GaAs/InAs Core/Shell Nanowires. *Nano Lett.* **2017**, *17*, 128–135.

(15) Basarić, F.; Wang, K.; Dumitru, T. G.; Manolescu, A.; Cesar, F. A.; Sanchez, A. M.; Krause, C.; Grützmacher, D.; Pawlis, A.; Schäpers, T. Interplay between Aharonov-Bohm and Altshuler-Aronov-Spivak Oscillations in Phase-Pure GaAs/InAs Core/Shell Nanowires of Different Lengths. *arXiv*:2606.10788.

(16) Gül, Ö.; Demarina, N.; Blömers, C.; Rieger, T.; Lüth, H.; Lepsa, M. I.; Grützmacher, D.; Schäpers, T. Flux Periodic Magnetoconductance Oscillations in GaAs/InAs Core/Shell Nanowires. *Phys. Rev. B* **2014**, *89*, 045417.

(17) Basarić, F.; Brajović, V.; Behner, G.; Moors, K.; Schaarman, W.; Juluri, R.; Sanchez, A. M.; Bae, J. H.; Lüth, H.; Grützmacher, D.; Pawlis, A.; Schäpers, T. Aharonov-Bohm and Altshuler-Aronov-Spivak Oscillations in the Quasiballistic Regime in Phase-Pure GaAs/InAs Core/Shell Nanowires. *Phys. Rev. B* **2025**, *112*, 075301.

(18) Zellekens, P.; Deacon, R. S.; Basaric, F.; Juluri, R.; Randle, M. D.; Bennemann, B.; Krause, C.; Zimmermann, E.; Sanchez, A. M.; Grützmacher, D.; Pawlis, A.; Ishibashi, K.; Schäpers, T. Flux-Periodic Supercurrent Oscillations in an Aharonov–Bohm-Type Nanowire Josephson Junction. *Commun. Phys.* **2025**, *8*, 363.

(19) Vaitiekėnas, S.; Winkler, G. W.; van Heck, B.; Karzig, T.; Deng, M. T.; Flensberg, K.; Glazman, L. I.; Nayak, C.; Krogstrup, P.; Lutchyn R. M.; Marcus C. M.; Shtrikman H. Flux-Induced Topological Superconductivity in Full-Shell Nanowires. *Science* **2020**, *367*, eaav3392.

(20) Payá, C.; Matute-Cañadas, F. J.; Yeyati, A. L.; Aguado, R.; San-Jose, P.; Prada, E. Fluxoid valve effect in full-shell nanowire Josephson junctions. *Phys. Rev. B* **2025**, *112*, 134520.

(21) Gül, Ö. Phase-coherent transport in GaAs/InAs core-shell nanowires. Master’s thesis, Julich Research Centre PGl-9, **2012**.

(22) Song, W.; Yu, Z.; Wang, Y.; Gao, Y.; Li, Z.; Yang, S.; Zhang, S.; Geng, Z.; Li, R.; Wang, Z.; Chen, F.; Yang, L.; Miao, W.; Xu, J.; Feng, X.; Wang, T.; Zang, Y.; Li, L.; Shang, R.; Xue, Q. K.; He K.; Zhang, H. Reducing Disorder in PbTe Nanowires for Majorana Research. *Nano Lett.* **2025**,

*25*, 2350-2356.

(23) Li, L.; Pan, D.; Xue, Y.; Wang, X.; Lin, M.; Su, D.; Zhang, Q.; Yu, X.; So, H.; Wei, D.; Sun, B.; Tan, P.; Pan, A.; Zhao, J. Near Full-Composition-Range High-Quality $GaAs_{1-x}Sb_x$ Nanowires Grown by Molecular-Beam Epitaxy. *Nano Lett.* **2017**, *17*, 622–630.

(24) Li, L.; Pan, D.; So, H.; Wang, X.; Yu, Z.; Zhao, J. GaAsSb/InAs Core-Shell Nanowires Grown by Molecular-Beam Epitaxy. *J. Alloys Compd.* **2017**, *724*, 659–665.

(25) Gül, Ö.; van Woerkom, D. J.; van Weperen, I.; Car, D.; Plissard, S. R.; Bakkers, E. P. A. M.; Kouwenhoven, L. P. Towards high mobility InSb nanowire devices. *Nanotechnology* **2015**, *26*, 215202.

(26) Geng, Z.; Zhang, Z.; Chen, F.; Yang, S.; Jiang, Y.; Gao, Y.; Tong, B.; Song, W.; Miao, W.; Li, R.; Wang, Y.; Zhang, Q.; Meng, F.; Gu, L.; Zhu, K.; Zang, Y.; Li, L.; Shang, R.; Feng, X.; Xue, Q. K.; He, K.; Zhang, H. Observation of Aharonov-Bohm effect in PbTe nanowire networks. *Phys. Rev. B* **2022**, *105*, L241112.

## Figure captions

**Fig. 1. Growth procedure and morphological characterization of $GaAs_{1-x}Sb_x$/InAs core-shell nanowires**. (a) Schematic illustration of the growth procedure: first, Ga self-catalyzed ternary $GaAs_{1-x}Sb_x$ nanowires are grown on Si (111) substrates; subsequently, the Ga droplets at the nanowire tips are removed; finally, InAs shells are epitaxially deposited on the sidewalls of the $GaAs_{1-x}Sb_x$ nanowire cores. The Sb content is varied to tune the lattice mismatch between $GaAs_{1-x}Sb_x$ and InAs. (b)-(d) Side-view SEM images of the core-shell nanowires with Sb contents $x$ = 0.15, 0.30, and 0.35, respectively. (e)-(g) Tilted-view (25°) SEM images of the corresponding nanowires for $x$ = 0.15, 0.30, and 0.35, respectively.

**Fig. 2. Microstructural characterization and defect density analysis of $GaAs_{1-x}Sb_x$/InAs core-shell nanowires with different Sb contents.** (a), (b) Low-magnification TEM images of the nanowires with Sb contents $x$ = 0.15 and 0.35, respectively. The coloured boxes in (a) and (b) delineate the areas where high-resolution TEM images were acquired. (c) Stacking fault density as a function of Sb content $x$ in the core-shell nanowires. (d), (f), (h), (j) High-resolution TEM images acquired from different areas of the $GaAs_{0.85}Sb_{0.15}$/InAs nanowire ($x$ = 0.15); (e), (g), (i), (k) the corresponding fast Fourier Transform (FFT) patterns. (l), (n), (p), (r) High-resolution TEM images from distinct regions of the $GaAs_{0.65}Sb_{0.35}$/InAs nanowire ($x$ = 0.35); (m), (o), (q), (s) the corresponding FFT patterns. The yellow arrow in (n) indicates the position of stacking faults.

**Fig. 3. Crystal structure and chemical composition of a cross-sectional sample of a $GaAs_{0.65}Sb_{0.35}$/InAs core-shell nanowire.** (a) Low-resolution cross-sectional TEM image of the entire cross-sectional specimen. (b)-(d) High-resolution TEM images of $GaAs_{0.65}Sb_{0.35}$, InAs, and their interface, respectively. The boxes labeled b, c, and d in (a) correspond to the regions where the high-resolution TEM images were recorded. (e)-(f) False-color EDS elemental maps of Ga and In of the sample, respectively. (g) False-color EDS elemental map of Ga and In (overlay) of the sample. (h) EDS line-scan profiles of Ga and In across the sample.

**Fig. 4. Mobilities and threshold voltages of InAs nanowires with a $GaAs_{0.85}Sb_{0.15}$ or a $GaAs_{0.65}Sb_{0.35}$ core**. (a) SEM of a $GaAs_{0.65}Sb_{0.35}$-InAs nanowire (device A) and the corresponding cross-sectional schematic. The scale bar is 400 nm. (b) Current-voltage *I-V* curves of device A measured at back gate voltages $V_{BG}$ = 5.05 V ~ 49.95 V with a gate step of 5 V. (c) Conductance ($G$) curves of device A (red) and device B ($x$ = 0.15, blue) as a function of $V_{BG}$. The arrow indicates the sweeping direction of $V_{BG}$. The dashed lines are fitting using the formula described in the main text. (d) $G$ as a function of $V_{BG}$ for all four devices. (e) Summary of the extracted mobilities: the red bars represent the two nanowires with $x$ = 0.35, whereas the blue bars represent those with $x$ = 0.15. Error bars represent mobility differences between the two devices. Upward

and downward indicate the sweeping directions of $V_{BG}$. (f) Summary of the threshold voltages ($V_{th}$).

**Fig. 5. Aharonov-Bohm-like oscillations in device A.** (a) Differential conductance ($dI/dV$) as a function of magnetic field $B$, applied parallel to the nanowire. $V_{BG}$ = 33.5V. $V$ = 0 mV. The gray curve shows the smoothed background (smoothing window ~ 0.4 T). The vertical lines are separated by 0.4 T. The $B$ range spans from -1.796 T to 1.796 T. (b) Oscillation signal after subtracting the smoothed background. (c) Fast Fourier Transform (FFT) amplitude of the trace in (b). A minor smoothing (averaged over 3 neighboring data points) was performed. The green dashed line indicates the dominant FFT peak at 2.5 $T^{-1}$. (d) Aharonov-Bohm oscillations at different back-gate voltages with background conductance subtracted. Adjacent traces are offset vertically by $0.1 \times 2e^2/h$ for clarity. From the top curve to the bottom one, $V_{BG}$ = 38 V, 36.5 V, 36 V, 35.25 V, 35 V, 34.5 V, 34.25 V, 34 V, 33.5 V, 33 V, 32.75 V and 31.75 V, respectively. (e) Averaged FFT amplitudes for all curves in (d). The red region indicates the standard deviation. A dominant FFT peak near 2.5 $T^{-1}$ is indicated by the green dashed line.

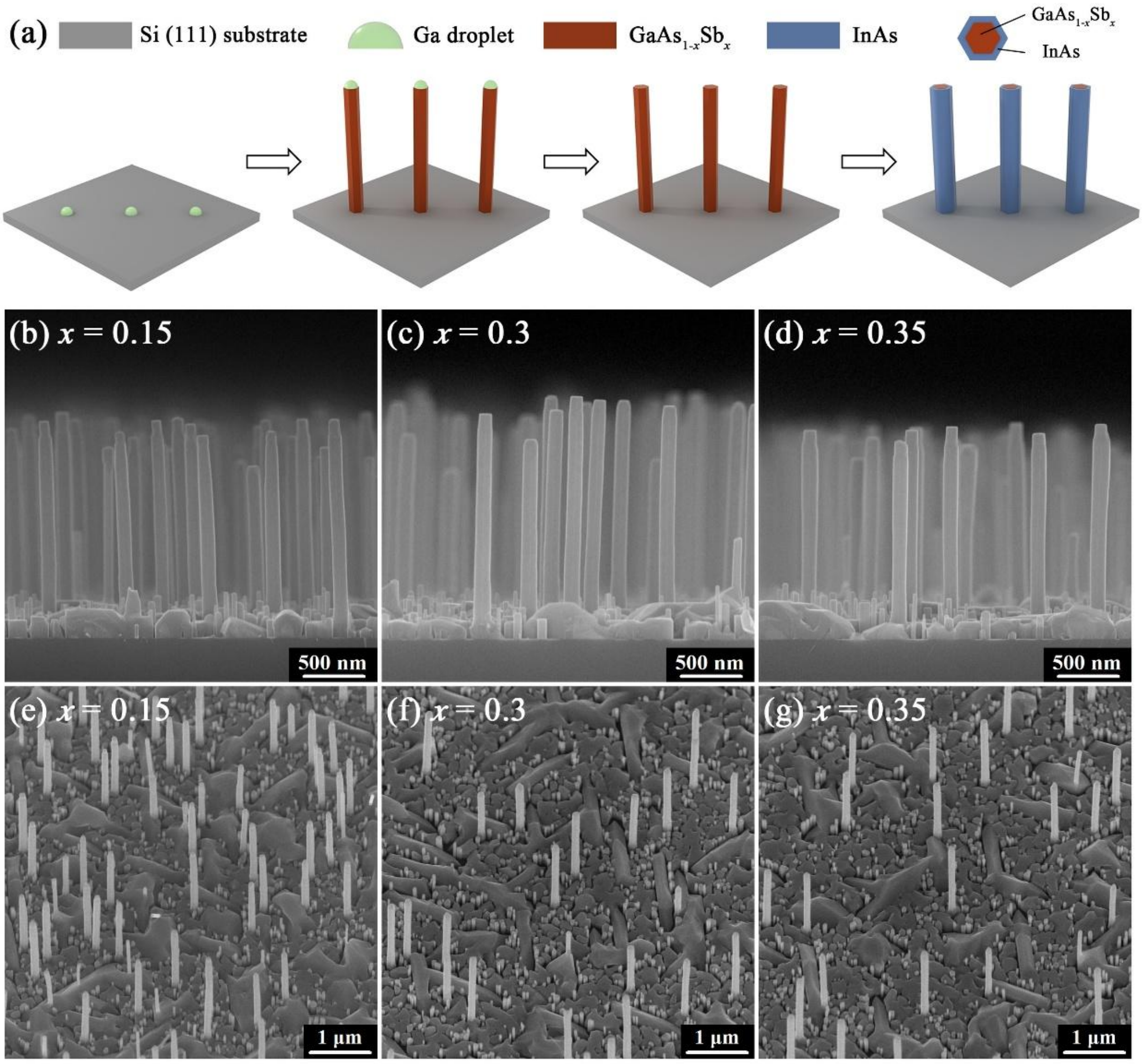
(a)
Si (111) substrate
Ga droplet
GaAs1-xSbx
InAs
GaAs1-xSbx
InAs
(b) x = 0.15
(c) x = 0.3
(d) x = 0.35
500 nm
500 nm
500 nm
(e) x = 0.15
(f) x = 0.3
(g) x = 0.35
1 μm
1 μm
1 μm

**Figure 1**

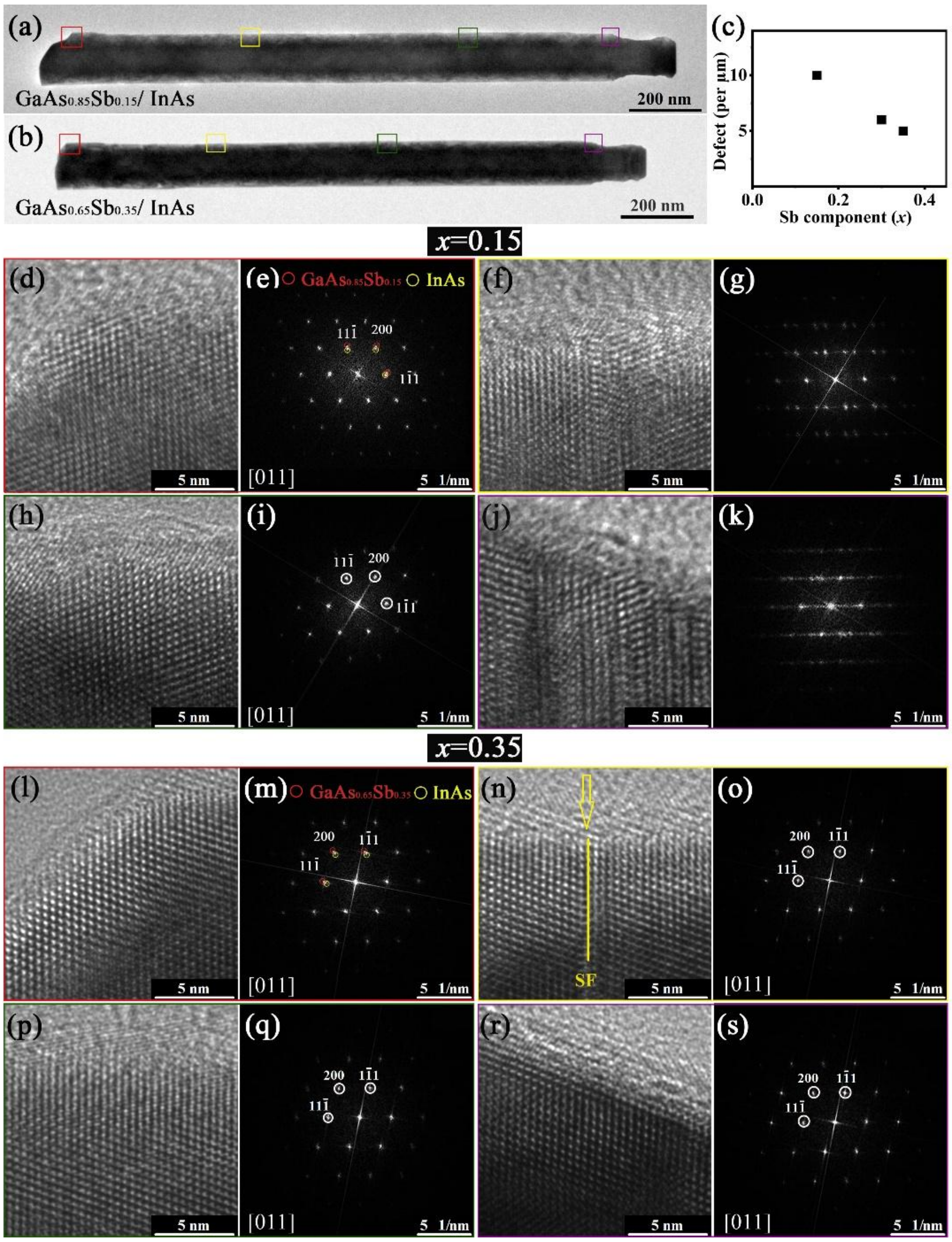
(a)
GaAs0.85Sb0.15/ InAs
200 nm
(b)
GaAs0.65Sb0.35/ InAs
200 nm
(c)
Defect (per μm)
Sb component (x)
x=0.15
(d)
(e) GaAs0.85Sb0.15 InAs
(f)
(g)
(h)
(i)
(j)
(k)
x=0.35
(l)
(m) GaAs0.65Sb0.35 InAs
(n)
SF
(o)
(p)
(q)
(r)
(s)
[011]
5 nm
5 1/nm

**Figure 2**

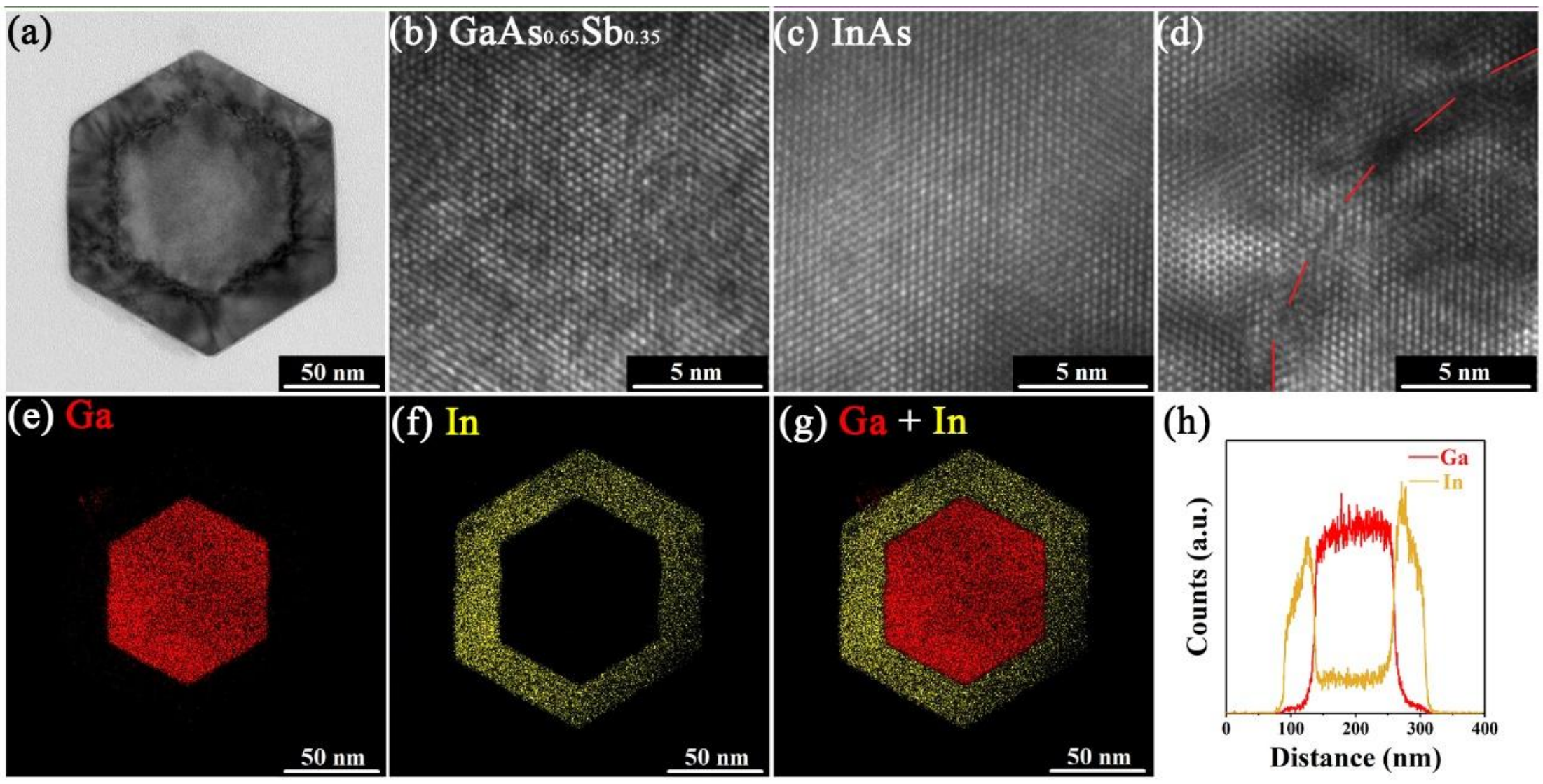
(a)
50 nm
(b) GaAs0.65Sb0.35
5 nm
(c) InAs
5 nm
(d)
5 nm
(e) Ga
50 nm
(f) In
50 nm
(g) Ga + In
50 nm
(h)
Ga
In
Counts (a.u.)
0
100
200
300
400
Distance (nm)

**Figure 3**

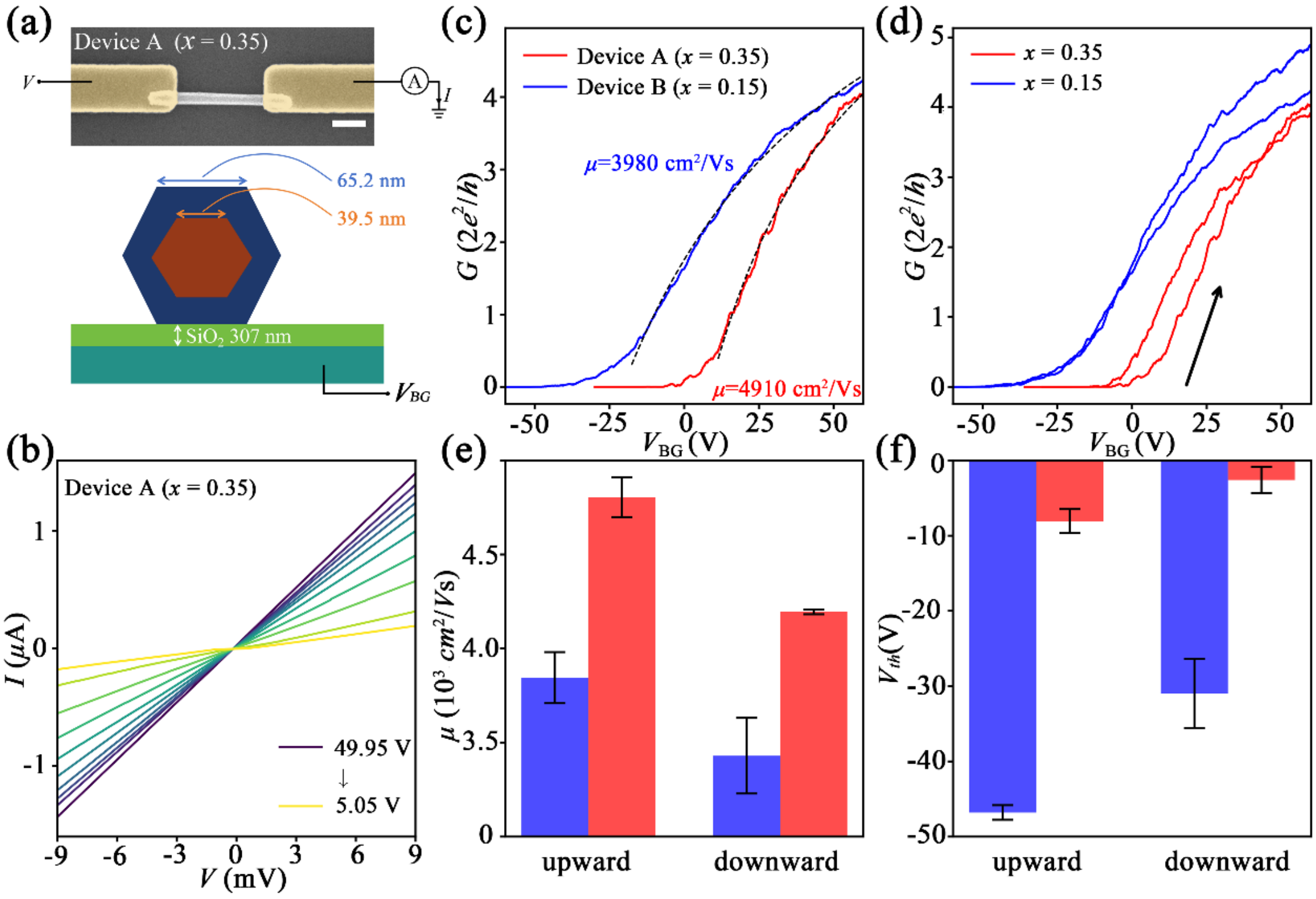
(a)
Device A (x = 0.35)
V
I
65.2 nm
39.5 nm
SiO2 307 nm
VBG
(b)
Device A (x = 0.35)
I (μA)
49.95 V
5.05 V
V (mV)
(c)
Device A (x = 0.35)
Device B (x = 0.15)
μ=3980 cm²/Vs
μ=4910 cm²/Vs
G (2e²/h)
VBG (V)
(d)
x = 0.35
x = 0.15
G (2e²/h)
VBG (V)
(e)
μ (10³ cm²/Vs)
upward
downward
(f)
Vth (V)
upward
downward

**Figure 4**

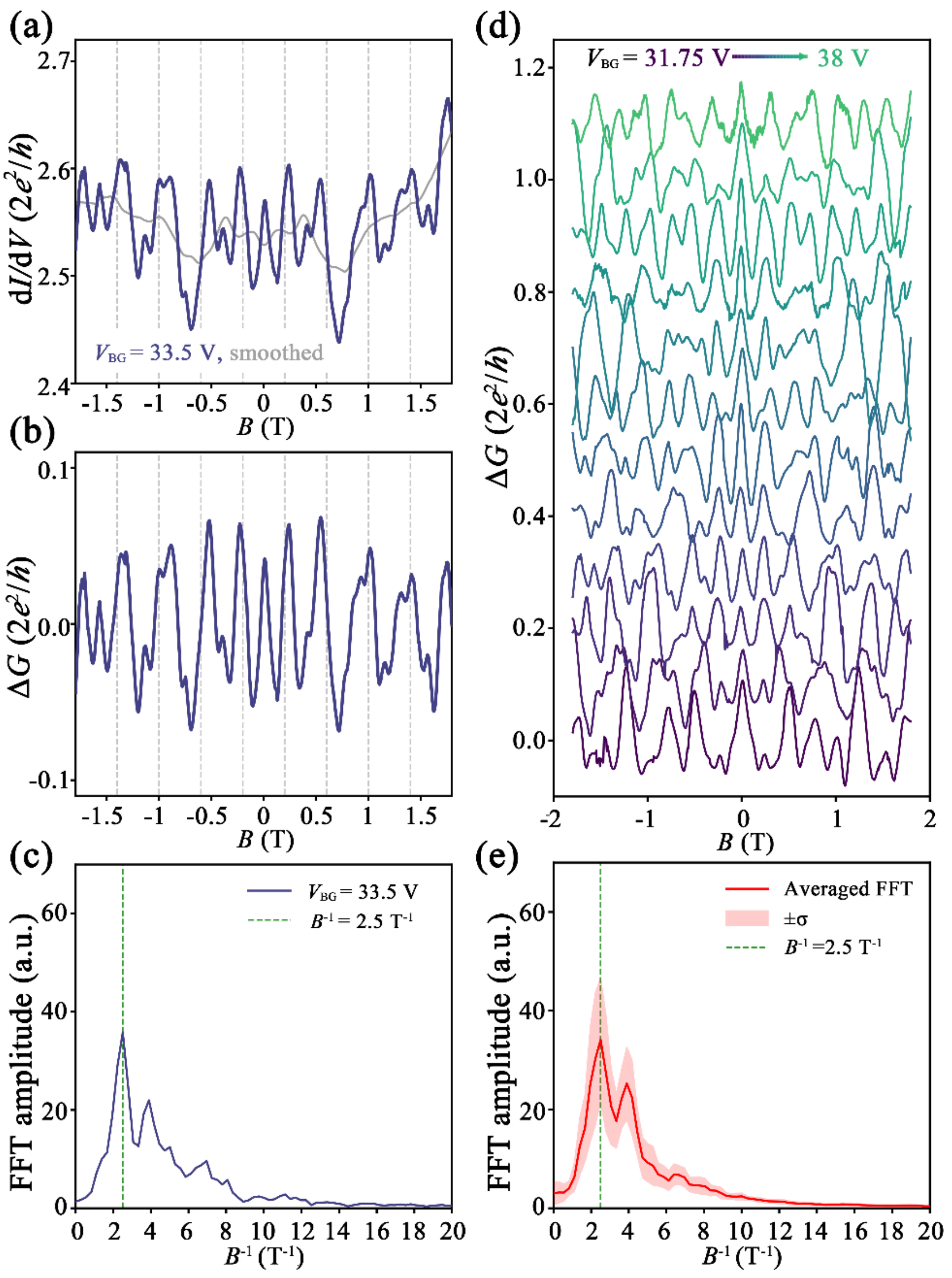
(a)
(b)
(c)
(d)
(e)
$V_{BG}$ = 33.5 V, smoothed
$V_{BG}$ = 31.75 V → 38 V
dI/dV (2e²/h)
ΔG (2e²/h)
B (T)
FFT amplitude (a.u.)
$V_{BG}$ = 33.5 V
B⁻¹ = 2.5 T⁻¹
Averaged FFT
±σ
B⁻¹ (T⁻¹)

**Figure 5**

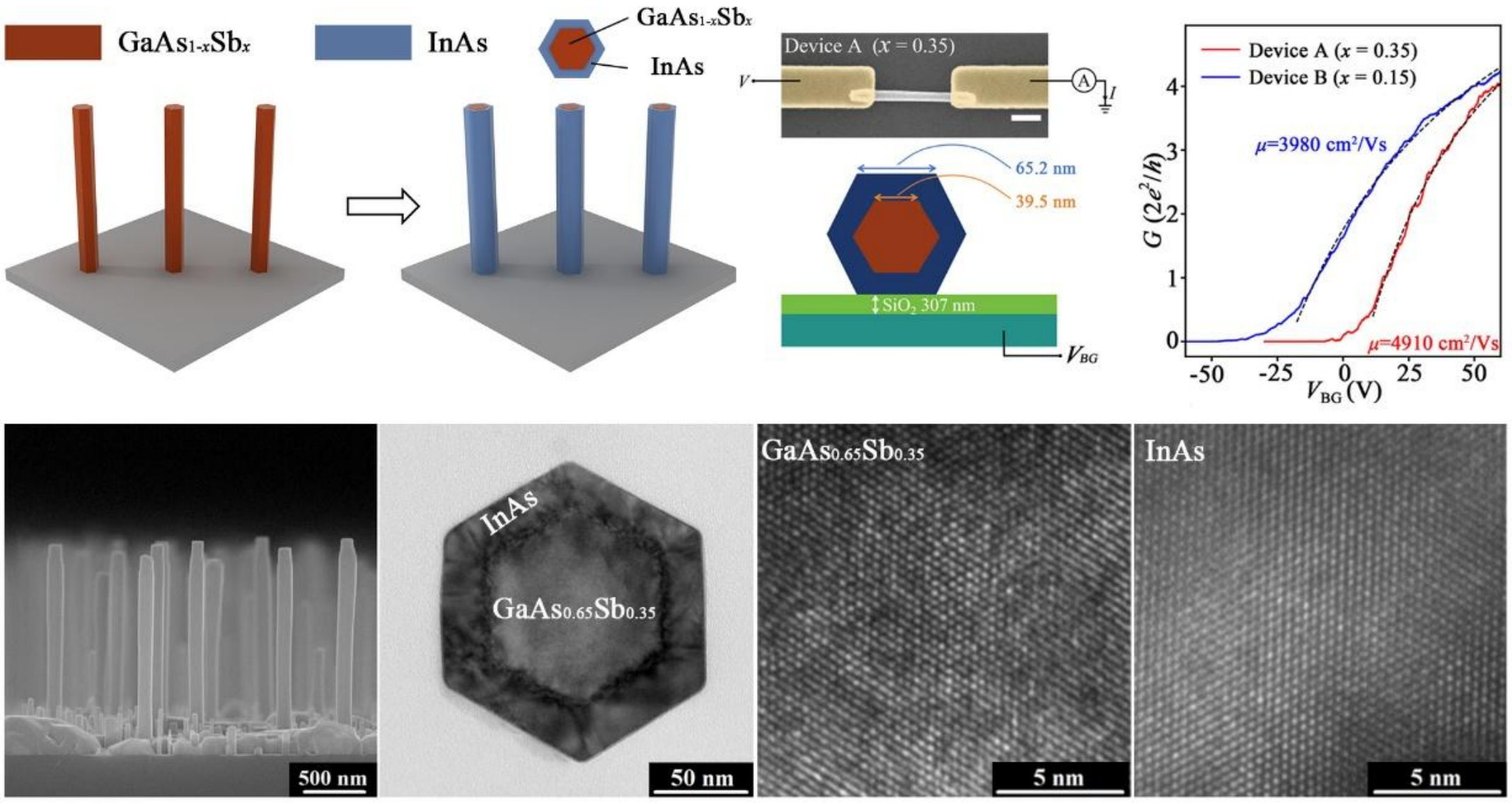

GaAs1-xSbx
InAs
GaAs1-xSbx
InAs
Device A (x = 0.35)
V
A
I
65.2 nm
39.5 nm
SiO2 307 nm
VBG
Device A (x = 0.35)
Device B (x = 0.15)
μ=3980 cm²/Vs
μ=4910 cm²/Vs
G (2e²/h)
VBG (V)
500 nm
InAs
GaAs0.65Sb0.35
50 nm
GaAs0.65Sb0.35
5 nm
InAs
5 nm

**TOC**

**Supporting Information**

# Controlled Growth and Disorder Mitigation of $GaAs_{1-x}Sb_x$/InAs Core-Shell Quantum Wires

Ran Zhuo[†,⊥,#], Huading Song[‡,#], Yanlin Wu[†,⊥,#], Wenlong Yang[†], Haiyan Shi[†], Xiyu Hou[†,⊥], Sijun Chen[§], Zhichuan Wang[||], Donghai Li[†,⊥], Jinjin Bai[†], Runan Shang[‡], Hao Zhang[§,||*], Dong Pan[†,⊥,*]

[†]*State Key Laboratory of Semiconductor Physics and Chip Technologies, Institute of Semiconductors, Chinese Academy of Sciences, Beijing 100083, China*

[⊥]*Center of Materials Science and Optoelectronics Engineering, University of Chinese Academy of Sciences, Beijing 100049, China*

[‡]*Beijing Academy of Quantum Information Sciences, Beijing 100193, China*

[§]*College of Semiconductors, Southern University of Science and Technology, Shenzhen 518055, China*

[||]*Quantum Science Center of Guangdong-Hong Kong-Macao Greater Bay Area, Shenzhen 518045, China*

[*]To whom correspondence should be addressed. E-mails: zhanghao@sustech.edu.cn; pandong@semi.ac.cn

[#]These authors contributed equally to this work.

**Statistical information on the diameter, length, and shell thickness of the $GaAs_{1-x}Sb_x$/InAs core-shell nanowires**

We have performed TEM measurements to obtain statistical information on the diameter, length, and shell thickness of the core-shell nanowires. The results are presented in Figs. S1-S3 for the three samples with different Sb contents. As shown in the figures, the nanowire length, core diameter, and shell thickness all exhibit a certain degree of distribution. Taking the $GaAs_{0.85}Sb_{0.15}$/InAs sample as an example (Fig. S1), the nanowire lengths of the ten representative nanowires range from approximately 1526 to 1758 nm, the core diameters range from approximately 75.4 to 85.5 nm, and the shell thicknesses range from approximately 13.1 to 22.6 nm. For the $GaAs_{0.65}Sb_{0.35}$/InAs sample (Fig. S3), the corresponding values are approximately 1694-1876 nm for the length, 101.9-122.8 nm for the core diameter, and 15.6-22.8 nm for the shell thickness. The intermediate sample with $x = 0.30$ (Fig. S2) shows similar distributions. Across the three samples, no systematic dependence of the nanowire length on the Sb content is evident, whereas both the core diameter and the shell thickness increase slightly with increasing Sb content, consistent with the larger total nanowire diameter observed at higher Sb content.

The observed distributions in diameter and shell thickness originate from the Ga self-catalyzed growth mode of the $GaAs_{1-x}Sb_x$ nanowires. Because the Ga droplet size is not uniform across the substrate owing to Ostwald ripening occurs during growth, the resulting $GaAs_{1-x}Sb_x$ core diameters exhibit a certain distribution. This, in turn, leads to a distribution in the InAs shell thickness. The data shown here represent the statistical information for the three samples, with 30 nanowires measured for each Sb content and 10 representative nanowires displayed in each figure.

Finally, we note that because the dimensions of individual nanowires differ, accurate extraction of transport parameters such as mobility requires precise knowledge of the core diameter and shell thickness for each measured device. Therefore, after completing the transport measurements, we performed FIB cross-sectioning on all devices and used TEM to determine the exact $GaAs_{1-x}Sb_x$ core diameter and InAs shell thickness, which were then used to calculate all transport data presented in the manuscript. This procedure ensures the reliability and reproducibility of the reported mobility values.

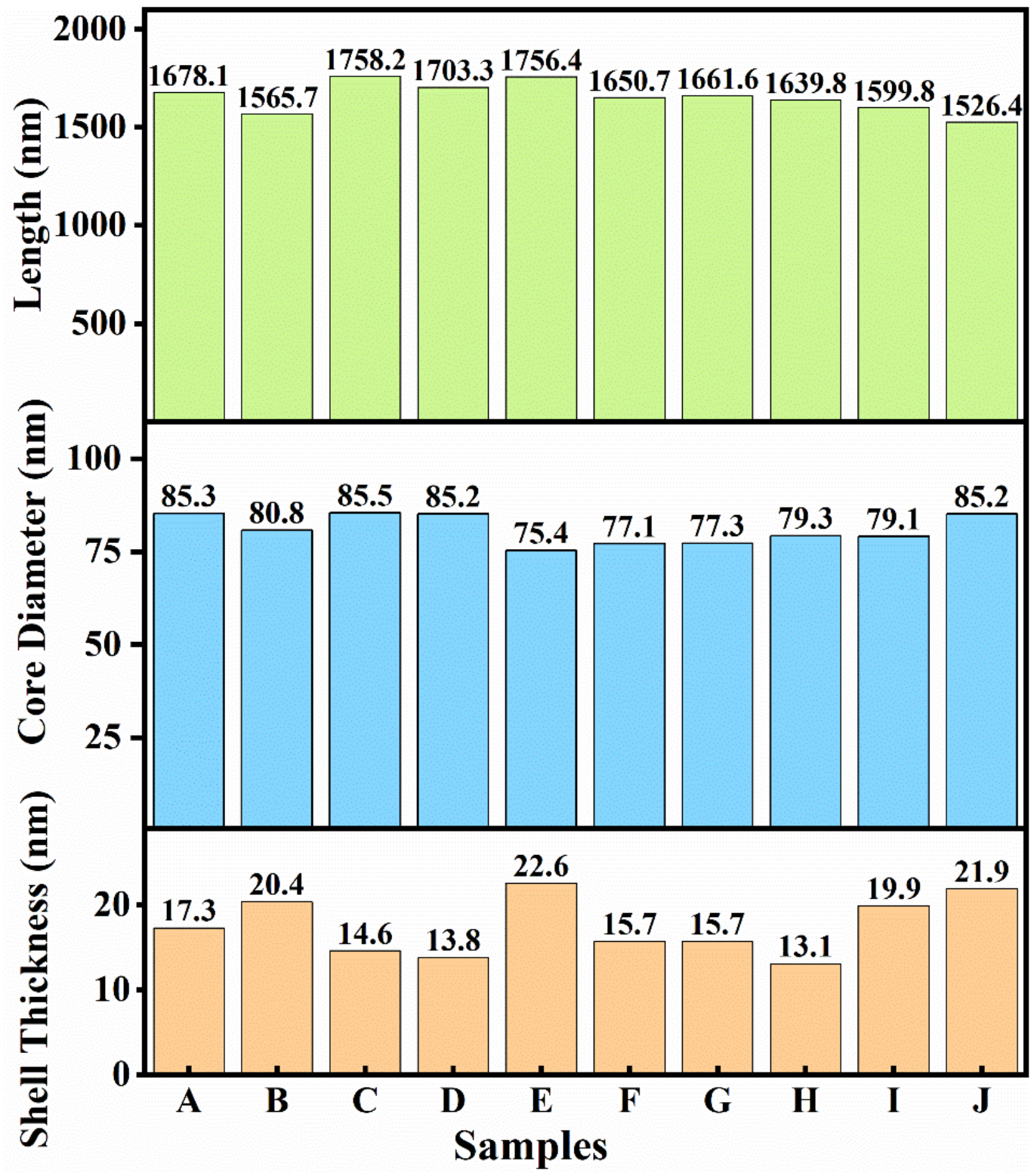

**Fig. S1.** Statistics of the length, diameter, and shell thickness of $GaAs_{0.85}Sb_{0.15}/InAs$ core-shell nanowires. The figure presents the statistical data obtained from ten representative nanowires.

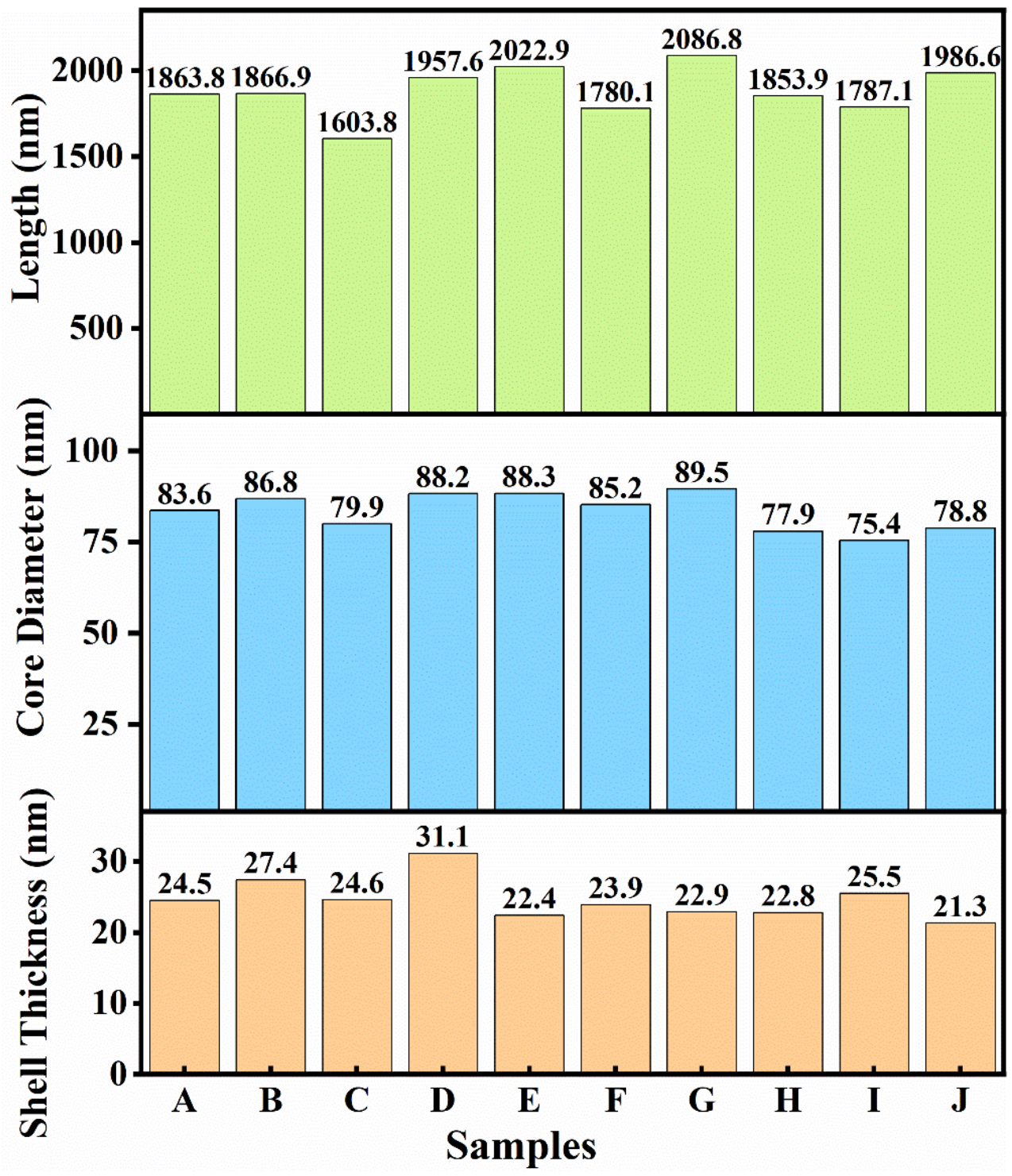

**Fig. S2.** Statistics of the length, diameter, and shell thickness of $GaAs_{0.70}Sb_{0.30}$/InAs core-shell nanowires. The figure presents the statistical data obtained from ten representative nanowires.

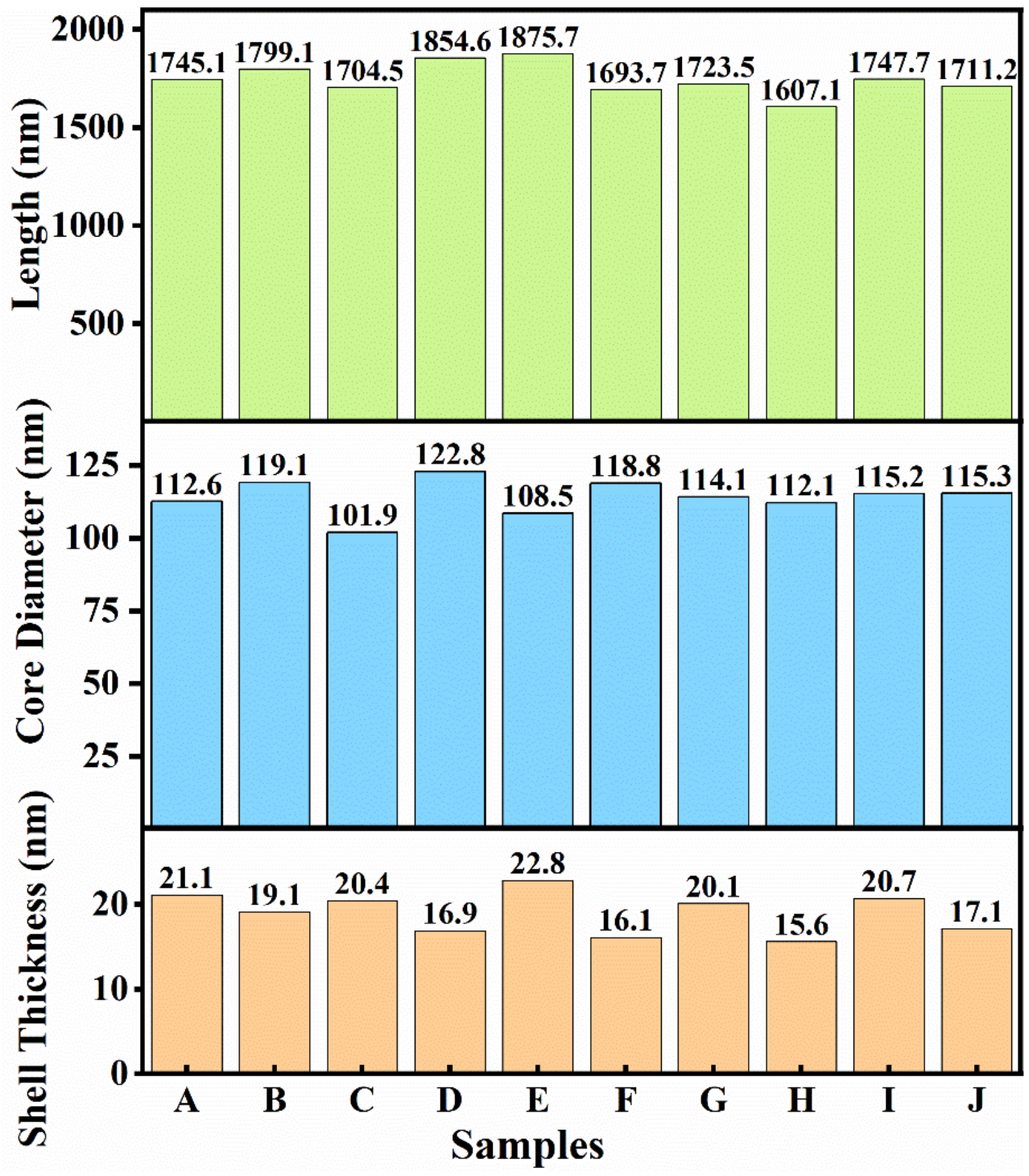

**Fig. S3.** Statistics of the length, diameter, and shell thickness of $GaAs_{0.65}Sb_{0.35}$/InAs core-shell nanowires. The figure presents the statistical data obtained from ten representative nanowires.

**Microstructural characterization and defect density analysis of $GaAs_{0.70}Sb_{0.30}$/InAs core-shell nanowires**

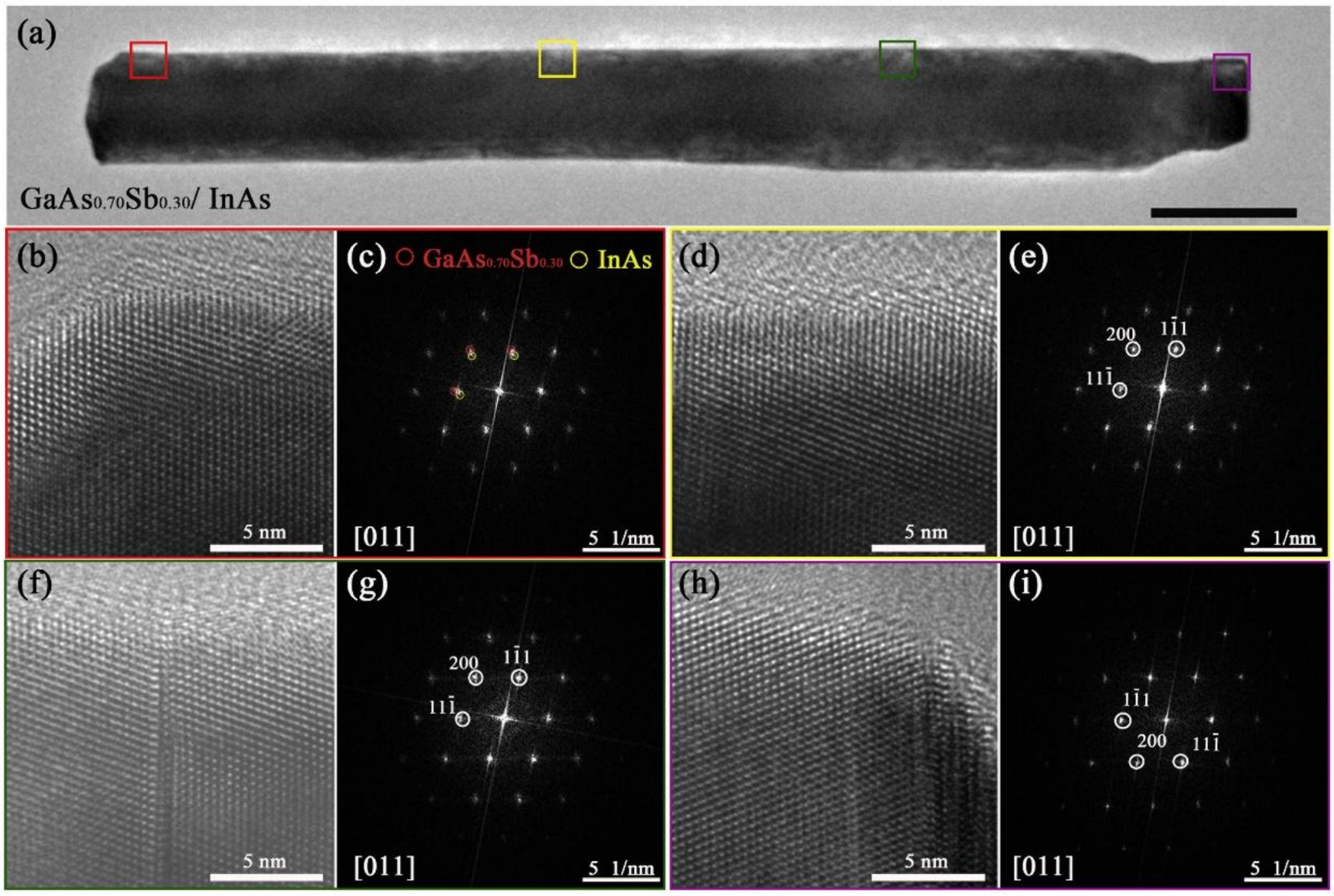

**Fig. S4. Microstructural characterization and defect density analysis of $GaAs_{0.70}Sb_{0.30}$/InAs core-shell nanowires.** (a) Low-magnification TEM image of the $GaAs_{0.70}Sb_{0.30}$/InAs core-shell nanowire. The coloured boxes delineate the specific areas where high-resolution TEM images were acquired. (b), (d), (f), (h) High-resolution TEM images acquired from different areas of the $GaAs_{0.70}Sb_{0.30}$/InAs nanowire; (c), (e), (g), (i) the corresponding fast Fourier Transform patterns. $GaAs_{0.70}Sb_{0.30}$/InAs core-shell nanowires also exhibit a zinc-blende crystal structure, with the InAs shell showing a defect density of approximately 6 $\mu m^{-1}$.

**Statistical analysis of the defect density in the $GaAs_{1-x}Sb_x$/InAs core-shell nanowires**

The defect density (stacking faults) in the nanowires was obtained from TEM measurements. We first examined the nanowires by low-magnification TEM, and then performed high-resolution TEM observations along the axial direction of each nanowire, from the bottom to the top, so that all defects present in the nanowire could be counted. For each Sb composition, more than 30 nanowires were measured, and the average value was taken as the defect density. Since the defects can be counted directly from the high-resolution TEM images, we did not provide an uncertainty value; the values we report are average values.

Figures S5 and S6 show typical TEM and high-resolution TEM images of $GaAs_{1-x}Sb_x$/InAs nanowires with Sb contents of 0.15 and 0.35, respectively. As seen in Fig. S5, for $x = 0.15$, 16 defects were found in a nanowire about 1.6 μm long, giving a stacking-fault density of approximately 10 $\mu m^{-1}$. Similarly, as seen in Fig. S6, for $x = 0.35$, 8 defects were found in a nanowire about 1.6 μm long, giving a stacking-fault density of approximately 5 $\mu m^{-1}$. These results show that increasing the Sb content from 0.15 to 0.35 halves the stacking-fault density in the InAs shell.

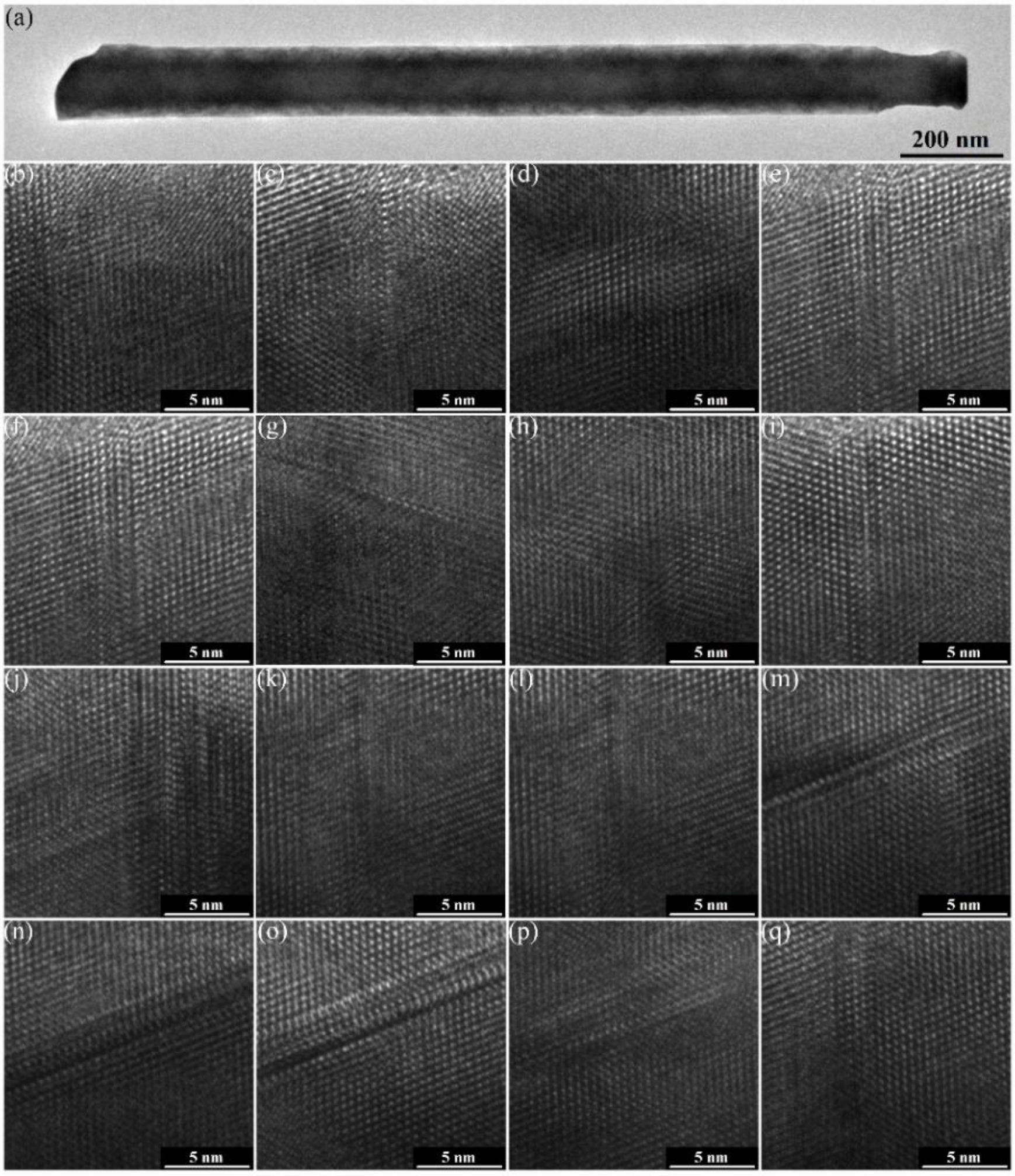

**Fig. S5.** TEM and high-resolution TEM images of $GaAs_{0.85}Sb_{0.15}/InAs$ core-shell nanowires. (a) Low-magnification TEM image of a typical nanowire; (b)-(q) High-resolution TEM images of all defect regions observed in the nanowire.

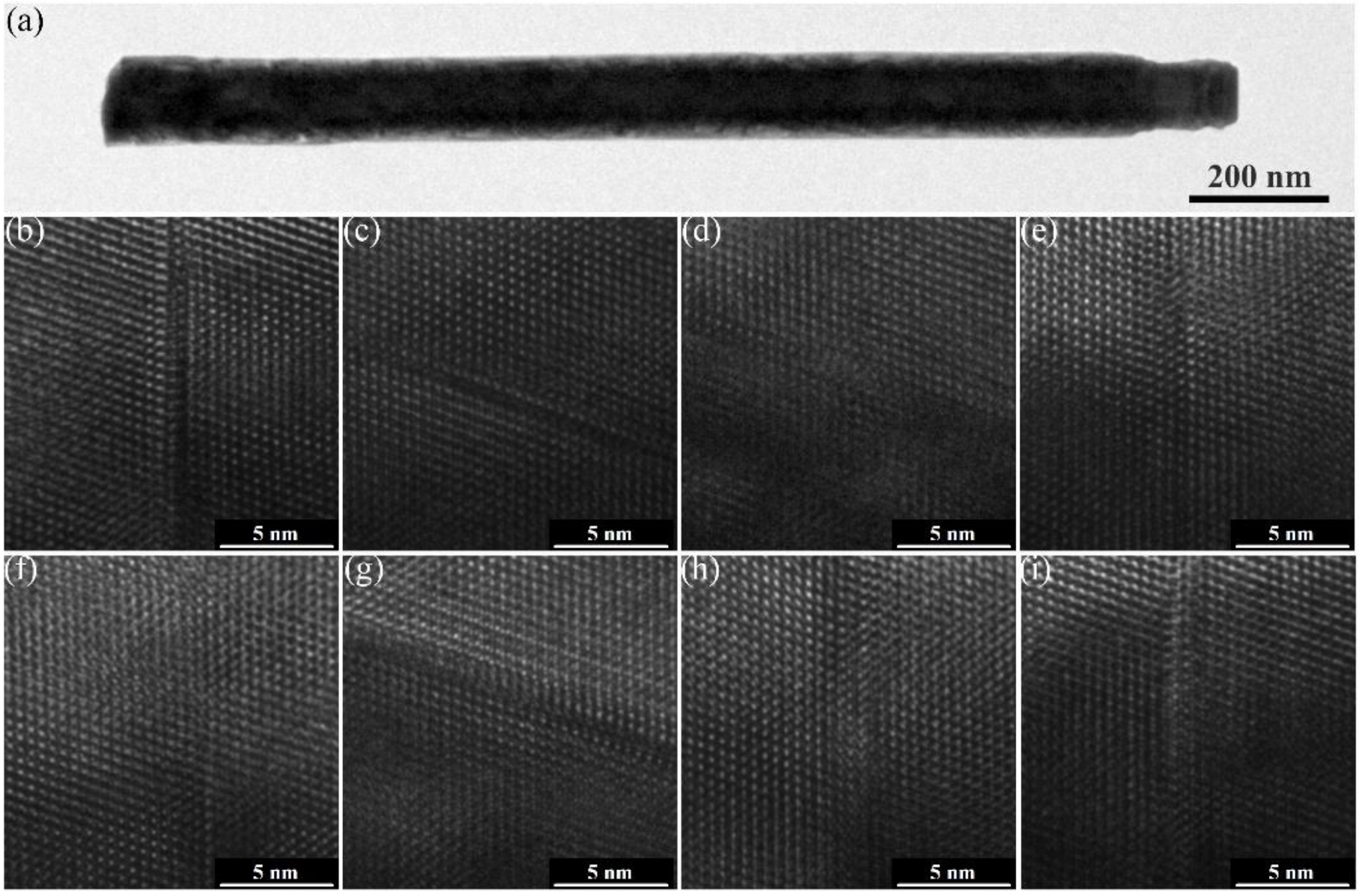

**Fig. S6.** TEM and high-resolution TEM images of $GaAs_{0.65}Sb_{0.35}/InAs$ core-shell nanowires. (a) Low-magnification TEM image of a typical nanowire; (b)-(i) High-resolution TEM images of all defect regions observed in the nanowire.

**Additional cross-sectional TEM characterization on the lower-Sb samples**

As mentioned above, in order to accurately extract the mobility and other transport parameters, we performed FIB cross-sectioning and TEM characterization on nanowire devices with all three Sb contents. Figs. S7 and S8 present the results for the samples with $x = 0.15$ and $x = 0.30$, respectively. Each figure includes (a) an SEM image of the device, (b) a low-magnification cross-sectional TEM image of the core-shell nanowire, and (c)-(f) high-resolution TEM images acquired at different locations of the core-shell interface. The white boxes in (b) highlight the regions from which the high-resolution TEM images were taken. During FIB specimen preparation, the lamella was cut along the radial direction of the nanowire, between the two metal electrodes.

As can be seen from the figures, for the low-Sb sample, the larger lattice mismatch gives rise to more pronounced defects at the core-shell interface. As the Sb content $x$ increases from 0.15 to 0.35, these defects become noticeably less frequent, consistent with the improved shell quality revealed by the plan-view TEM analysis in the main text. We note, however, that the defect distribution observed in a cross-sectional specimen depends on the specific cutting position. Therefore, to obtain a more reliable and statistically meaningful measure of the defect density, we used conventional plan-view TEM imaging in the main text to quantify the stacking-fault density, rather than relying on cross-sectional specimens alone. The cross-sectional results shown here provide additional, consistent evidence that increasing the Sb content improves the crystalline quality of the InAs shell.

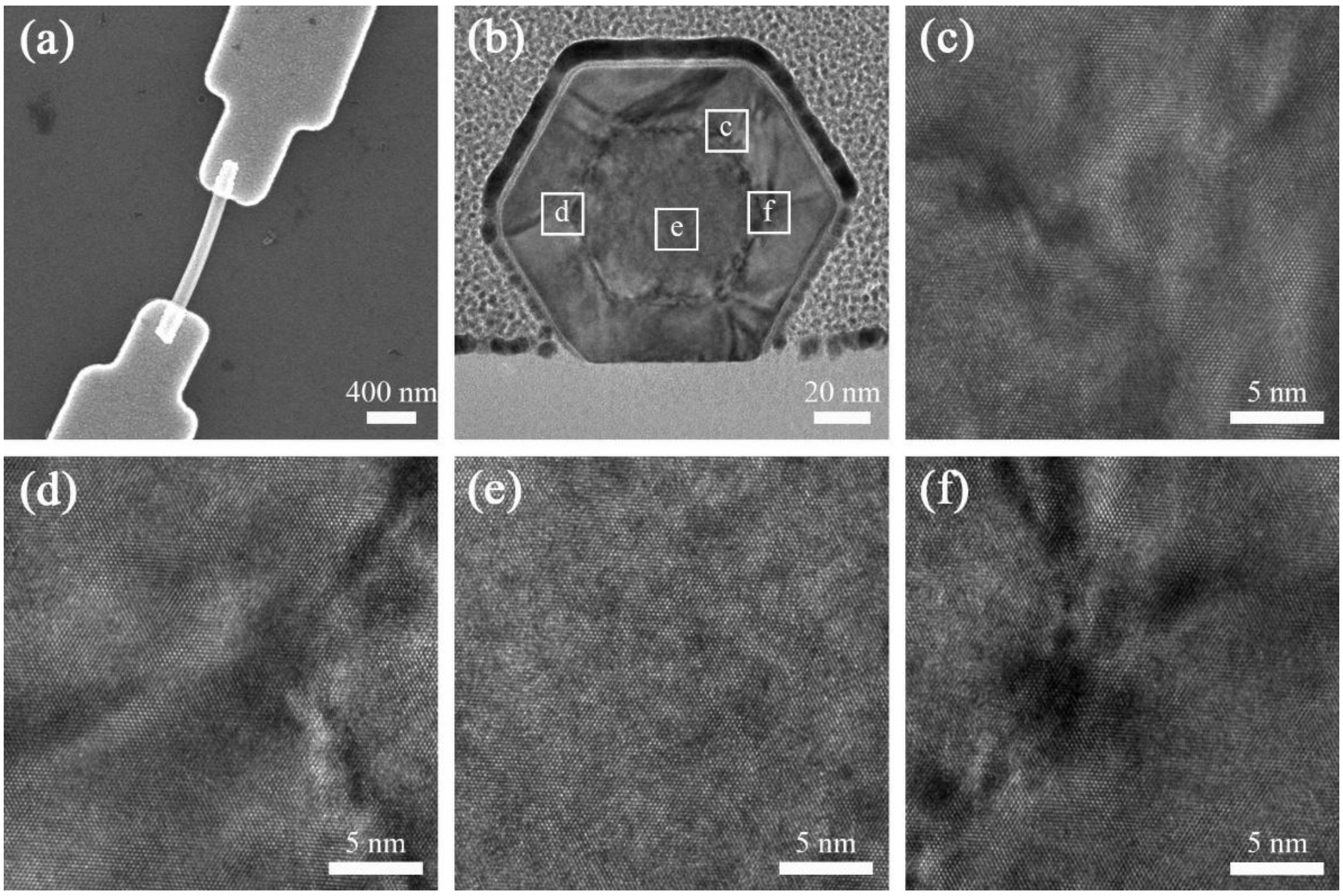

**Fig. S7.** (a) SEM of the $GaAs_{0.85}Sb_{0.15}$/InAs nanowire device. (b) Low-magnification cross-sectional TEM image of the core-shell nanowire. During FIB specimen preparation, the lamella was cut along the radial direction of the nanowire, between the two metal electrodes. (c)-(f) High-resolution TEM images acquired at different locations of the core-shell interface. The white boxes in (b) highlight the regions from which the high-resolution TEM images were taken.

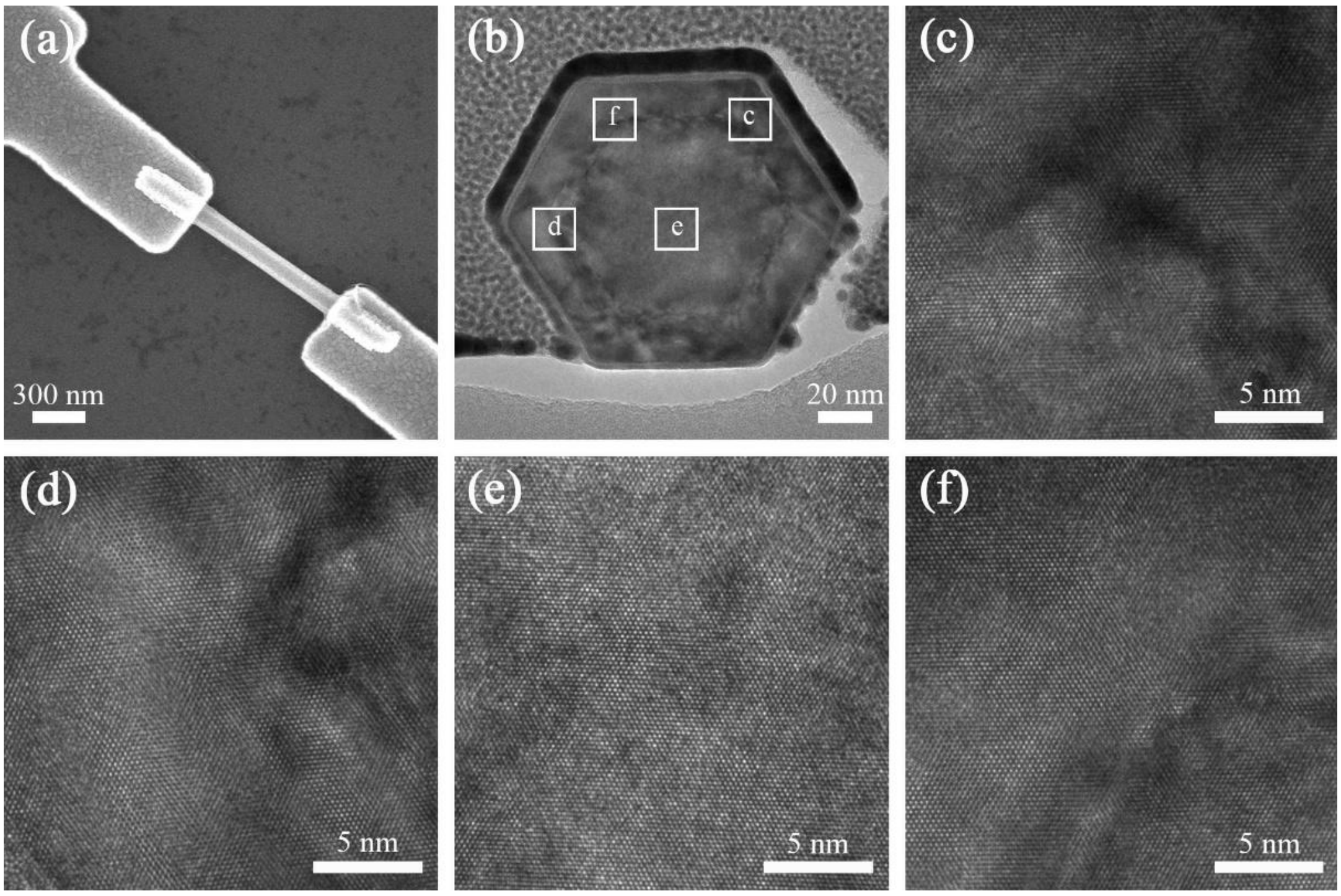

**Fig. S8.** (a) SEM of the $GaAs_{0.70}Sb_{0.30}$/InAs nanowire device. (b) Low-magnification cross-

sectional TEM image of the core-shell nanowire. During FIB specimen preparation, the lamella was cut along the radial direction of the nanowire, between the two metal electrodes. (c)-(f) High-resolution TEM images acquired at different locations of the core-shell interface. The white boxes in (b) highlight the regions from which the high-resolution TEM images were taken.

## Additional transport data of InAs nanowires epitaxially grown on ternary $GaAs_{1-x}Sb_x$ nanowires with different Sb contents

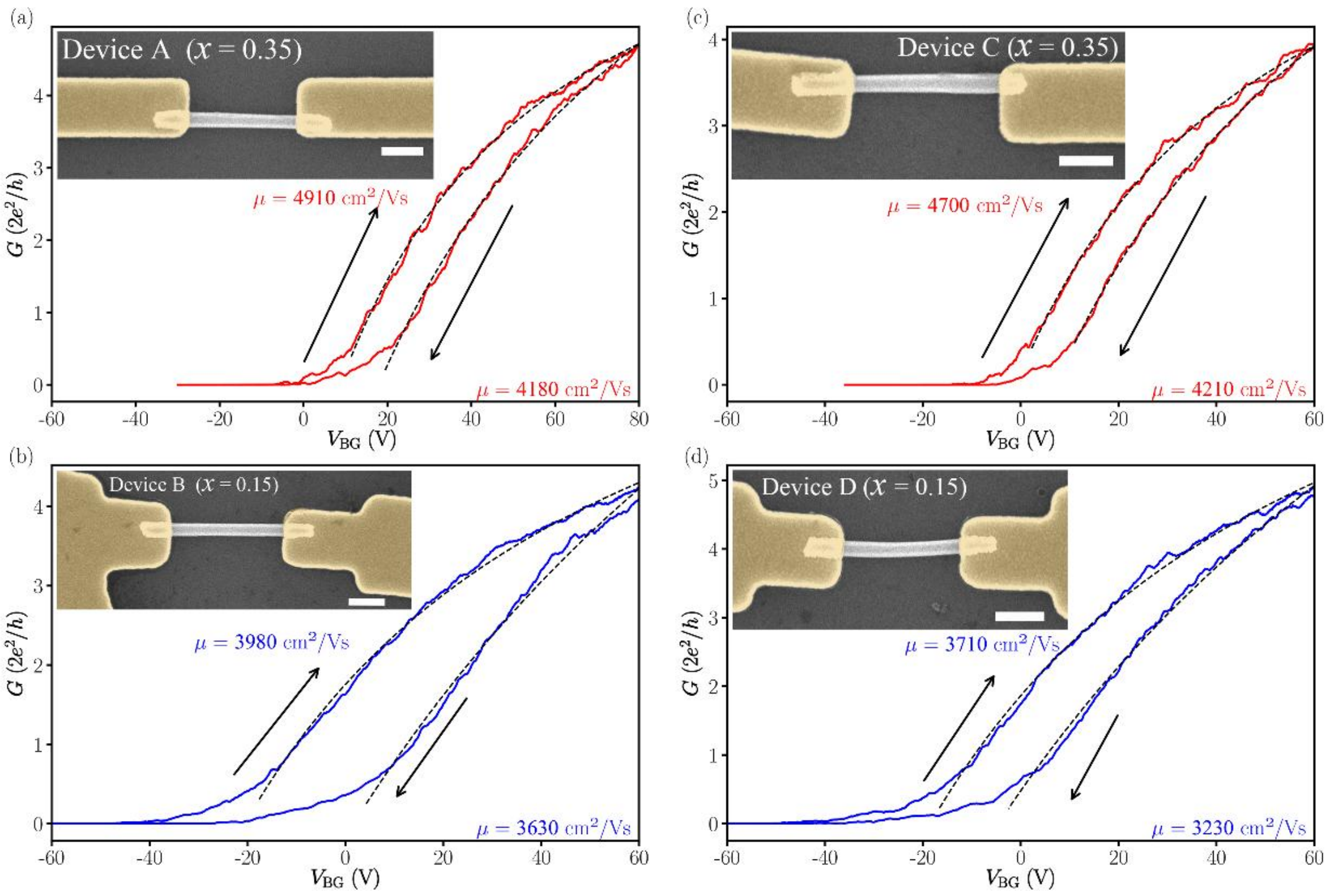

**Fig. S9.** (a-d) Pinch-off curves of both sweeping directions (arrows) of the four devices in Fig. 4. Insets are their SEM images. The dashed lines are mobility fits. Scale bars: 400 nm.

**Table S1.** The device parameters of the four devices in Fig. 4.

| Devices | $GaAs_{1-x}Sb_x$/InAs | Length (nm) | Radius (nm) | Capacitance (aF) | Mobility upward ($cm^2$/Vs) | Mobility downward ($cm^2$/Vs) | $V_{th}$ upward (V) | $V_{th}$ downward (V) |
|---|---|---|---|---|---|---|---|---|
| **A** | $x = 0.35$ | 1013 | 65.2 | 26.3 | 4910 | 4180 | -9.65 | -4.35 |
| **B** | $x = 0.15$ | 1246 | 67.1 | 32.4 | 3980 | 3630 | -47.78 | -35.55 |
| **C** | $x = 0.35$ | 1075 | 59.5 | 26.6 | 4700 | 4210 | -6.48 | -0.85 |
| **D** | $x = 0.15$ | 964 | 62.0 | 24.0 | 3710 | 3230 | -45.83 | -26.40 |

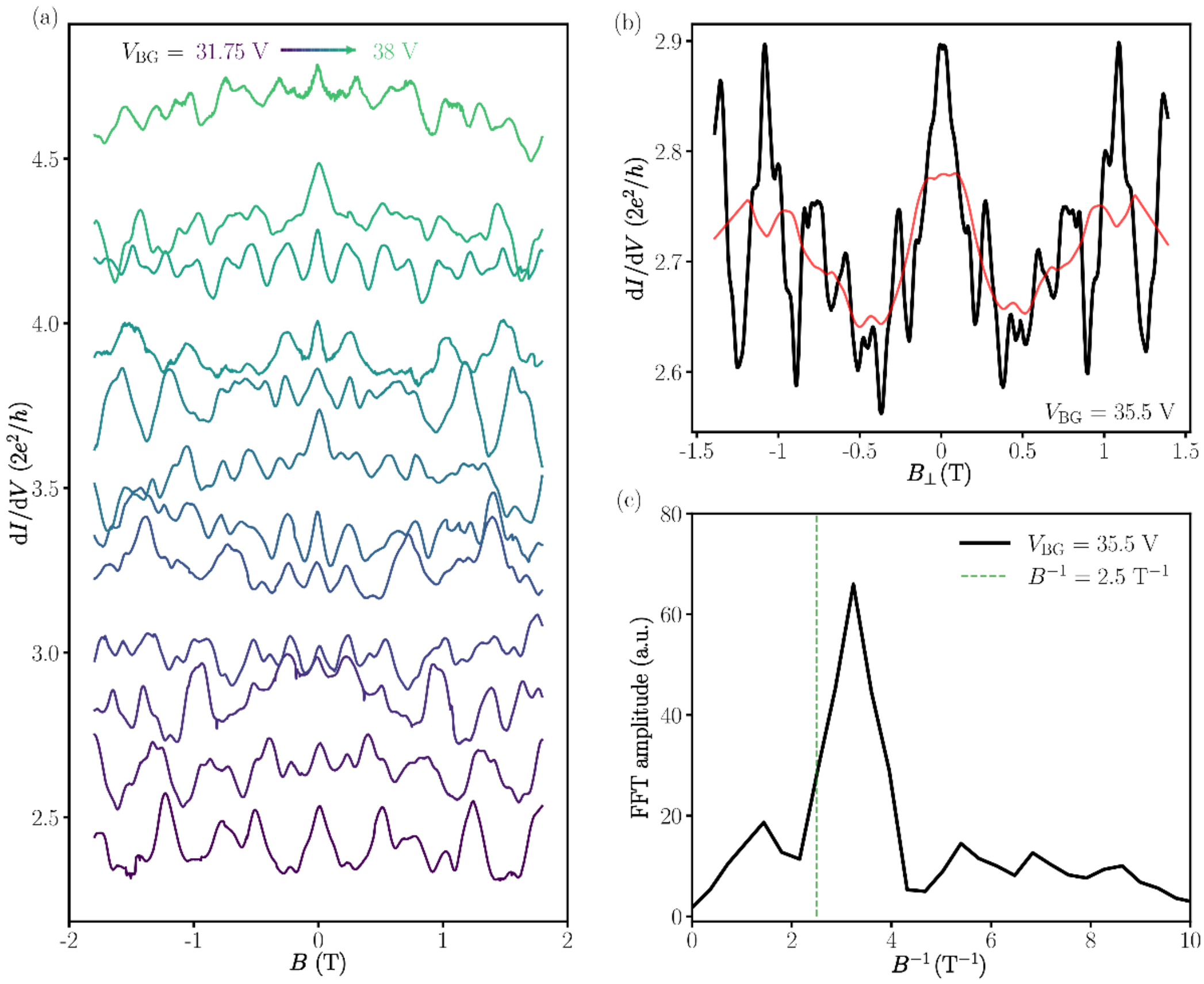

**Fig. S10.** (a) Differential conductance (d$I$/d$V$) curves of Fig. 5(d) before background subtraction. Vertical offsets are $0.15\times 2e^2/h$ for clarity. (b) Magneto-conductance by applying $B$ perpendicular to the device substrate. $V_{BG}=35.5$V, $V=0$ mV. The smoothed background of the original data is displayed by the red curve (smooth window ~ 0.4 T). (c) FFT amplitudes of (b), the FFT peak is significantly shifted from 2.5 T$^{-1}$ (green dashed line).

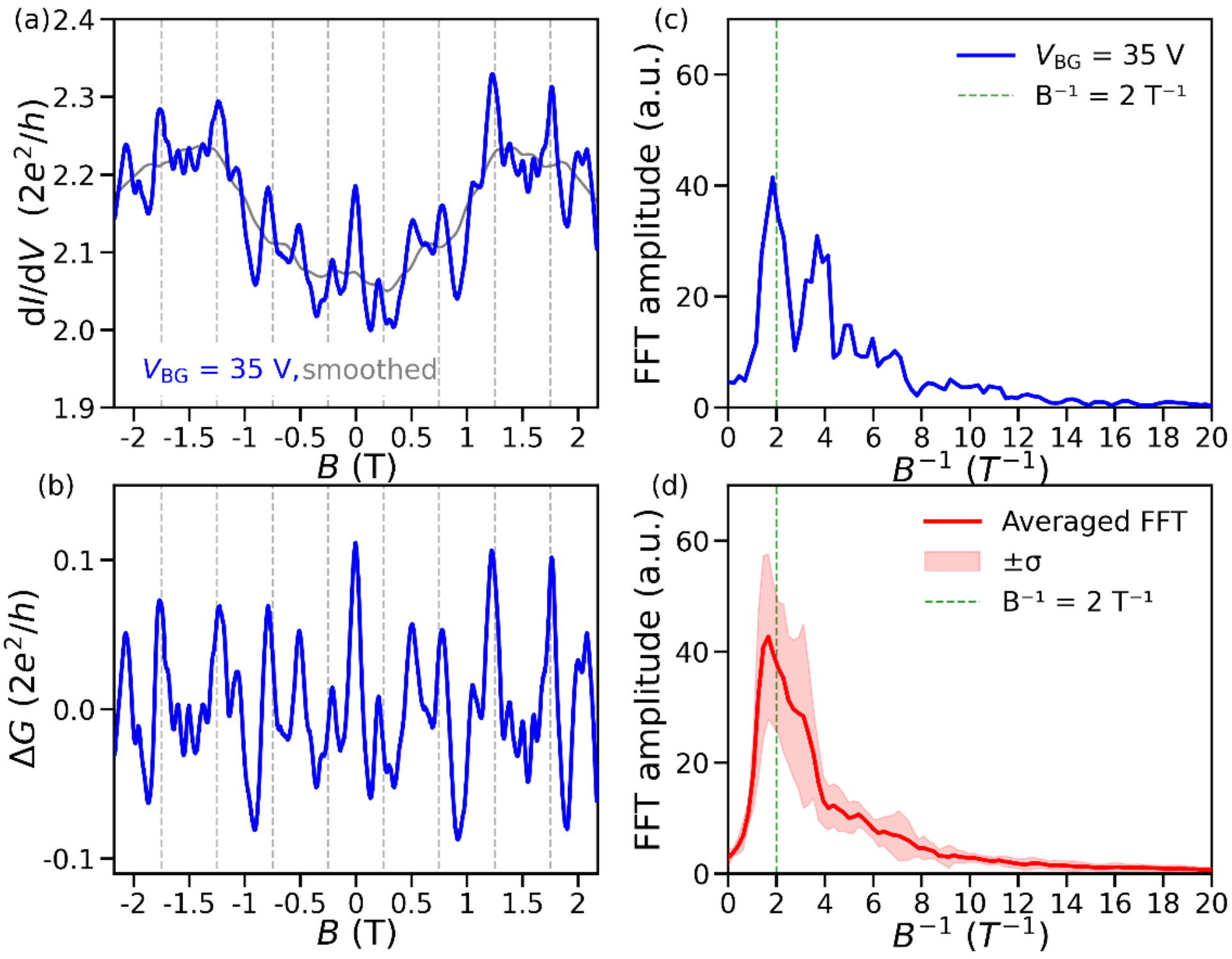

**Fig. S11.** Aharonov-Bohm-like oscillations in device C. (a) Differential conductance (d$I$/d$V$) as a function of magnetic field $B$, applied nearly parallel to the nanowire. $V_{BG}$ = 35 V and $V$ = 0 mV. The gray curve shows the smoothed background (smoothing window ~ 0.5 T). The vertical lines are separated by 0.5 T. The $B$ range spans from -2.176 T to 2.176 T. (b) Oscillation signal after subtracting the smoothed background. (c) FFT amplitude of the trace in (b). A minor smoothing (averaged over 3 neighboring data points) was performed. The green dashed line indicates the dominant FFT peak at 2 $T^{-1}$, corresponding to period of 0.5 T. (d) Averaged FFT amplitudes for all d$I$/d$V$ – $B$ curves of device C. The gate voltages $V_{BG}$ = 25 V, 30 V, 33 V, 35 V, 37.5 V, 38 V, 38.5 V, 39 V, 39.5 V and 40 V. The red region indicates the standard deviation. A dominant FFT peak near 2 $T^{-1}$ is indicated by the green dashed line.